\documentclass[12pt]{article}
\usepackage{amsfonts,amssymb,amsmath,graphicx,placeins,mathtools,amsthm}
\usepackage{subfigure}
\usepackage{nomencl}
\usepackage{chngcntr}
\usepackage[auth-lg,affil-it]{authblk}
\usepackage{array,booktabs}
\usepackage[table]{xcolor}
\usepackage[english]{babel} 
\usepackage{cite}
\usepackage{tabu} 
\usepackage[a4paper,left=25mm,right=25mm,top=25mm,bottom=30mm]{geometry}
\usepackage[auth-lg,affil-it]{authblk}
\begin{document}

\title{Rheological Model for Wood}
 
\author[a]{Mohammad Masoud Hassani}
\author[a]{Falk K.~Wittel}
\author[a]{Stefan Hering}
\author[a]{Hans J.~Herrmann}
\affil[a]{Computational Physics for Engineering Materials, IfB, ETH Zurich, Schafmattstrasse 6, CH-8093 Zurich, Switzerland}

\maketitle

\begin{abstract}
Wood as the most important natural and renewable building material plays an important role in the construction sector. Nevertheless, its hygroscopic character basically affects all related mechanical properties leading to degradation of material stiffness and strength over the service life. Accordingly, to attain reliable design of the timber structures, the influence of moisture evolution and the role of time- and moisture-dependent behaviors have to be taken into account. For this purpose, in the current study a 3D orthotropic elasto-plastic, visco-elastic, mechano-sorptive constitutive model for wood, with all material constants being defined as a function of moisture content, is presented. The corresponding numerical integration approach, with additive decomposition of the total strain is developed and implemented within the framework of the finite element method (FEM). Moreover to preserve a quadratic rate of asymptotic convergence the consistent tangent operator for the whole model is derived. 

Functionality and capability of the presented material model are evaluated by performing several numerical verification simulations of wood components under different combinations of mechanical loading and moisture variation. Additionally, the flexibility and universality of the introduced model to predict the mechanical behavior of different species are demonstrated by the analysis of a hybrid wood element. Furthermore, the proposed numerical approach is validated by comparisons of computational evaluations with experimental results. 

\emph{\bfseries{Keywords}}: Hardwood, Constitutive model, Moisture, Multi-surface plasticity, Numerical integration, Mechano-sorption, Moisture-stress analysis.

\end{abstract}
\section{Introduction}\label{Sec:Intro}
Wood application, in particular in form of engineered or composite wood elements, has significantly increased as structural building material. Only recently, structural wooden components such as glulam or laminated veneer lumber (LVL) entirely or partially (hybrid) out of hardwood aim for or obtained general technical approvals. The advantages of hardwood such as beech, oak, or ash are obvious: high strength and stiffness allow for smaller cross-section or span width compared to softwood, resulting in increased dimensional stability, load-carrying capacity and finally more freedom for design. In spite of the aforementioned advantages, unfortunately the strong hygric dependence of basically all mechanical properties render many innovative ideas futile. In addition, time-dependent phenomena like long-term visco-elastic creep \cite{Holzer89,Liu93} and mechano-sorption under changing environmental conditions \cite{Houska95,Bengtsson99,Hanhij2000} can accelerate degradation of stiffness and strength over the life-time of a structural wood component and result in the loss of capacity and consequently structural integrity even after being in use for decades. Realistic long-term predictions of the mechanical performance of hardwood or hybrid elements under external mechanical and climatic loading should be a central concern for assuring both - serviceability and safety of timber structures. It is a common practice today to disregard moisture, creep, plasticity, and mechano-sorption for technical approvals, not because they are insignificant, but because of difficulties in their experimental assessment, that is challenging due to a high degree of coupling, time consuming due to low diffusivity, and in some cases simply impossible. For these reasons and in order to attain effective design criteria for new products, the development and characterization of an authentic moisture-dependent constitutive material model for different wood species and the robust implementation of its corresponding rheological model in broadly used non-linear finite element (FE) environments is of great importance.

Non-linear numerical models for wood, in particular with long-term character are very rare due to the need of species and moisture-dependent mechanical parameters. Several experimental and theoretical constitutive wood material models with different ambient relative humidity (RH) and mechanical loading were proposed. A review on the general use of FE in wood analysis was published by Meckerle \cite{Mackerle05}, while other authors focus on reviewing proposed rheological models \cite{Hanhi95Thes,Chassagne07}. In principle the sources for non-linearity like plasticity, creep, or mechano-sorption are covered to quite different extent. Creep and mechano-sorption were addressed in \cite{SantTchRep91,Ormarsson99,HanHelnI2003,HanHelnII2003,Chassagne06,Fortino09,BorKalisEberh12}, while others focus on elasto-plastic behavior, disregarding time dependency \cite{HelnShell05,GuanZhu09,Oudjene09}. A comprehensive constitutive model comprising all potentially activated mechanical responses as a function of moisture content for different species is still missing. However it is the basis for numerical simulations that grant insight into the long-term behavior of structural elements of hardwoods, softwoods and hybrid ones.

Here a 3D orthotropic material model with moisture content is presented, where all instantaneous and time-dependent deformation mechanisms are considered. The corresponding numerical implementation of the material model is written as a material (UMAT) subroutine for the use within a finite element environment. The material model can be utilized as the basis of any moisture-stress analysis, for non-linear fracture mechanical problems established on the cohesive zone model and applications of interface elements, or de-bonding simulations under different combinations of service loadings and changing environmental conditions. The manuscript is organized as follows: First all partial deformations as components of the total strain together with their corresponding rheological formulations are described from a thermodynamical point of view. Then the description of moisture transfer inside wood elements and the principles of moisture-stress analysis are summarized. Section~\ref{Sec:NumInt} focuses on the iterative time integration approach and presents all theoretical formulations needed for the numerical implementation of the material model in the context of a finite element environment. The verification of the functionality and applicability as well as the experimental validation of the material model are shown in Section~\ref{Sec:Verif} by means of examples. Finally the generality and flexibility of the proposed constitutive model are demonstrated by a hygro-mechanical simulation for a typical hybrid glulam beam made of European beech and Norway spruce in Section~\ref{Sec:Example}. 
\section{Description of the hygro-mechanical constitutive model}\label{Sec:Model}
\begin{figure*}[htpb]
	  \begin{center}
	\includegraphics[scale=0.60]{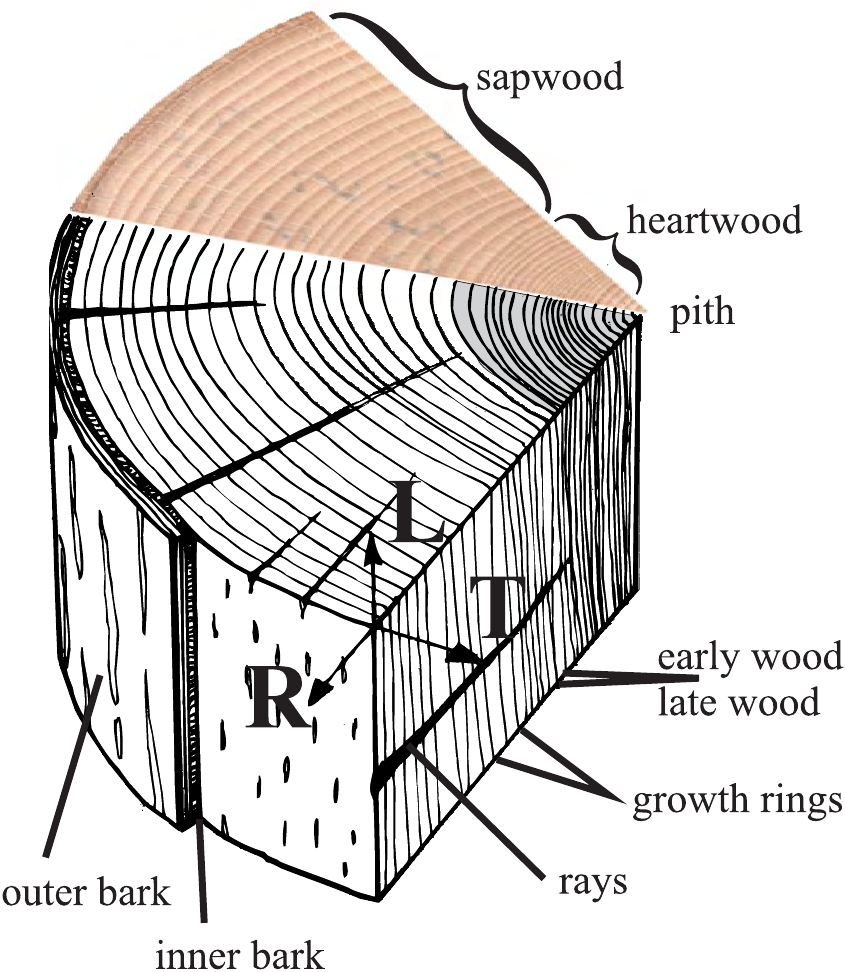}
	\caption{Definition of the local orthotropy directions of wood.}
	\label{fig:OrthoDir}
	  \end{center}
\end{figure*}

Due to the cellular nature of wood and its growth, wood is in general anisotropic. However for cut sections distant from the center of the stem, it is usually considered as an orthotropic material with three major axes, namely longitudinal along fiber direction, radial and tangential in the plane perpendicular to grain (see Fig.~\ref{fig:OrthoDir}). The curvature of the growth rings can be considered by defining the orthotropic material in a cylindrical coordinate system. To capture the consequences of the hygric behavior, all mechanical properties along the anatomical directions have to be consequently expressed as a function of moisture content.

All potentially participating deformation mechanisms have to be considered when moisture content changes during the use of wood structures. To predict the true deformation field and for the subsequent stress analysis, the respective material model consists of deformation modes originating from
\begin{itemize}
	\item elastic deformation $\boldsymbol{\varepsilon}^{el}$ (Section~\ref{SSSec:Elastic}),
	\item irrecoverable plastic deformation $\boldsymbol{\varepsilon}^{pl}$ (Section~\ref{SSSec:Plastic}),
	\item swelling/shrinkage called hygro-expansion in this context $\boldsymbol{\varepsilon}^{\omega}$ (Section~\ref{SSSec:Hygro}),
	\item visco-elastic creep $\boldsymbol{\varepsilon}_{i}^{ve}$ (\emph{Kelvin-Voigt} element-wise visco-elastic strain tensor) (Section~\ref{SSSec:Visco}),
	\item mechano-sorption $\boldsymbol{\varepsilon}_{j}^{ms}$ (\emph{Kelvin-Voigt} element-wise mechano-sorptive strain tensor) (Section~\ref{SSSec:Mechano}).
\end{itemize}
\begin{figure*}[htpb]
	  \begin{center}
	\includegraphics[scale=0.70]{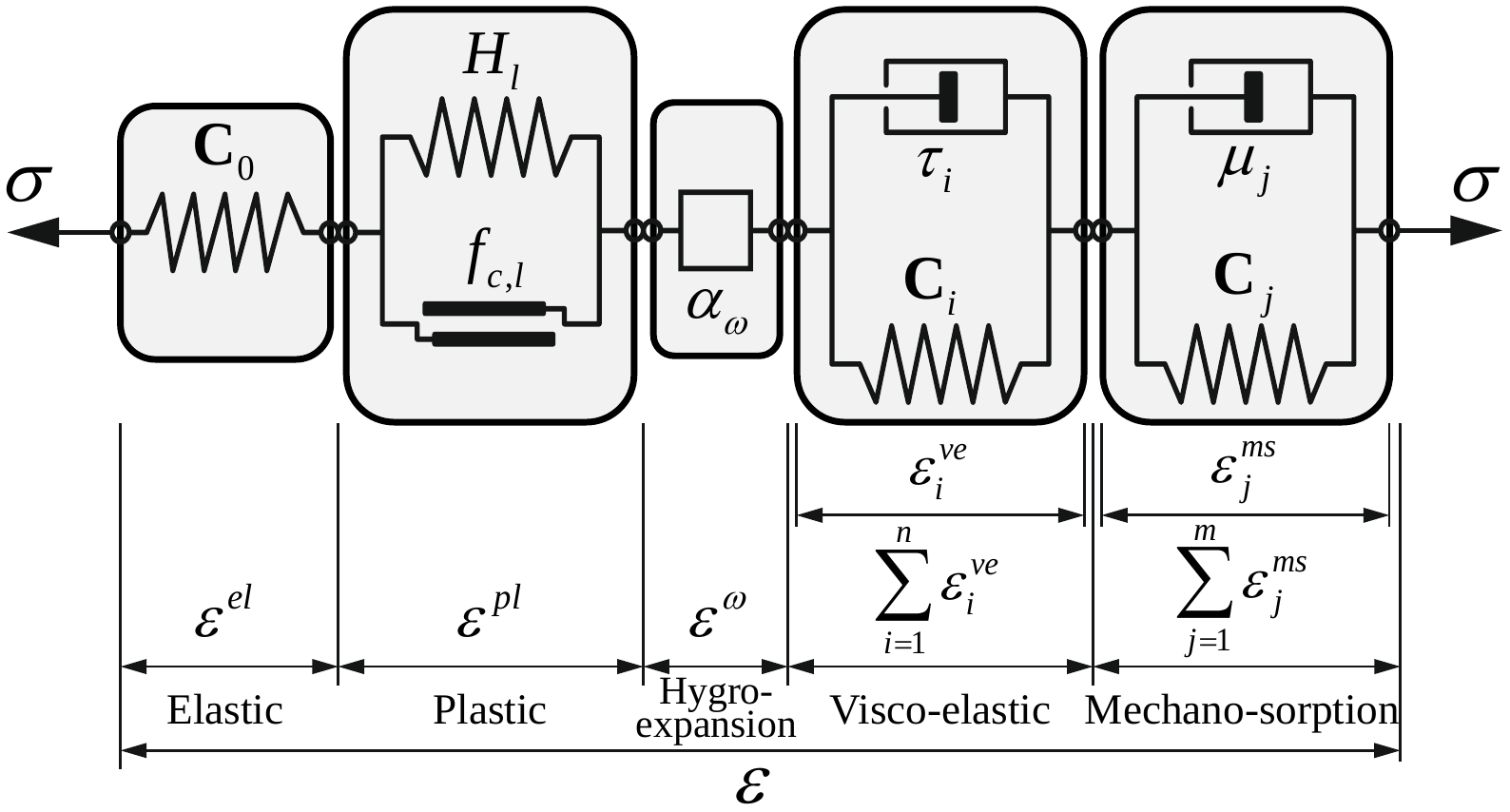}
	\caption{Schematic illustration of the constitutive material model.}
	\label{fig:MatModel}
	  \end{center}
\end{figure*}

In general a total strain tensor~$\boldsymbol{\varepsilon}$ is defined that accumulates the respective contributions in the way: (see \cite{HanHelnI2003,HanHelnII2003,Fortino09} and Fig.~\ref{fig:MatModel})
\begin{equation}  \label{eq:TotStrain}
\boldsymbol{\varepsilon}=\boldsymbol{\varepsilon}^{el}+\boldsymbol{\varepsilon}^{pl}+\boldsymbol{\varepsilon}^{\omega}+\sum_{i=1}^{n} \boldsymbol{\varepsilon}_{i}^{ve}+\sum_{j=1}^{m} \boldsymbol{\varepsilon}_{j}^{ms}.    
\end{equation}
From a thermodynamical perspective \cite{HanHelnI2003,HanHelnII2003,Fortino09}, the free energy function,~$\psi$, is defined as:
\begin{equation} \label{eq:Psi}
\psi =\psi(T,\omega,\boldsymbol{\varepsilon},\boldsymbol{\varepsilon}^{pl},\boldsymbol{\varepsilon}_{i}^{ve},\boldsymbol{\varepsilon}_{j}^{ms},\alpha_{l})=\phi(T,\omega)+\psi^{el}+\psi^{ve}+\psi^{ms}+\frac{1}{2}\sum_{l=1}^{r} q_{l}\alpha_{l}.
\end{equation}
Here~$\phi(T,\omega)$ is a general expression for the thermal energy and since in the present study the effect of temperature on the mechanical response of the material is ignored, it is not further considered.~$\psi^{el}$ specifies elastic strain energy,~$\psi^{ve}$ and~$\psi^{ms}$ represent energy accumulated in the visco-elastic and mechano-sorptive elements:
\begin{equation}     \label{eq:PsiDef}
\psi^{el} = \frac{1}{2}\boldsymbol{\varepsilon}^{el}:{\bf C}_{0}:\boldsymbol{\varepsilon}^{el},~~~~~~~~\psi^{ve} = \frac{1}{2}\sum_{i=1}^{n}\boldsymbol{\varepsilon}_{i}^{ve}:{\bf C}_{i}:\boldsymbol{\varepsilon}_{i}^{ve},~~~~~~~\psi^{ms} = \frac{1}{2}\sum_{j=1}^{m}\boldsymbol{\varepsilon}_{j}^{ms}:{\bf C}_{j}:\boldsymbol{\varepsilon}_{j}^{ms},
\end{equation}
with the elastic stiffness tensor~${\bf C}_{0}$, and the element-wise visco-elastic and mechano-sorptive stiffness tensors~${\bf C}_{i}$ and~${\bf C}_{j}$. Finally, the last term of the right-hand-side of Eq.~(\ref{eq:Psi}) designates the isotropic hardening energy, which appears during the evolution of irrecoverable plastic deformations. In the following, all partial strains based on an additive decomposition of the total strain in addition to their associated thermodynamic driving stresses are described following \cite{HanHelnI2003,HanHelnII2003}. It should be noted the proposed hygro-mechanical constitutive model is formulated and implemented for infinitesimal strains. It means the application of the material model in case of large deformations, non-linear visco-elasticity, or damage phenomenon is not relevant.
\nomenclature{$\boldsymbol{\varepsilon}/\boldsymbol{\varepsilon}^{el}/\boldsymbol{\varepsilon}^{pl}/\boldsymbol{\varepsilon}^{\omega}$}{Total strain/Elastic strain/Plastic strain/Hygro-expansion strain tensor}
\nomenclature{$\boldsymbol{\varepsilon}_{i/j}^{ve/ms}$}{$i/j^{th}$ visco-elastic/mechano-sorptive strain tensor}
\nomenclature{$n/m$}{Number of visco-elastic/mechano-sorptive \emph{Kelvin-Voigt} elements}
\nomenclature{$\psi/\phi(T,\omega)$}{Free energy/Thermal energy function}
\nomenclature{$\psi^{el/ve/ms}$}{Elastic/Visco-elastic/Mechano-sorptive strain energy}
\subsection{Description of mechanical behavior}\label{SSec:Mechbehav}
\subsubsection{Elastic deformation}\label{SSSec:Elastic}
The elastic deformation represents the scleronomous linear and fully recoverable material behavior. After differentiating the free energy function Eq.~(\ref{eq:Psi}) with respect to the total strain, the corresponding rheological relation is obtained as:
\begin{equation}     \label{eq:Elas}
\boldsymbol{\sigma}=\frac{\partial\psi}{\partial\boldsymbol{\varepsilon}}={\bf C}_{0}:\left(\boldsymbol{\varepsilon}-\boldsymbol{\varepsilon}^{pl}-\boldsymbol{\varepsilon}^{\omega}-\sum_{i=1}^{n} \boldsymbol{\varepsilon}_{i}^{ve}-\sum_{j=1}^{m} \boldsymbol{\varepsilon}_{j}^{ms}\right)={\bf C}_{0}:\boldsymbol{\varepsilon}^{el}.      
\end{equation}
Consequently one can write~$\boldsymbol{\varepsilon}^{el}={\bf C}_{0}^{-1}:\boldsymbol{\sigma}$, with the orthotropic elastic compliance tensor~${\bf C}_{0}^{-1}$ constituted by 9 independent material engineering constants given as
\renewcommand{\arraystretch}{1.5}
\begin{equation}
     {\bf C}_{0}^{-1}=\left[ \begin{array}{cccccc}
          \frac{1}{E_{R}}\;         & \;\frac{-\nu_{TR}}{E_{T}}\; & \;\frac{-\nu_{LR}}{E_{L}}\; & \;0\;                & \;0\; & \;0 \\
          \frac{-\nu_{RT}}{E_{R}}\; & \;\frac{1}{E_{T}}\;         & \;\frac{-\nu_{LT}}{E_{L}}\; & \;0\;                & \;0\; & \;0 \\
          \frac{-\nu_{RL}}{E_{R}}\; & \;\frac{-\nu_{TL}}{E_{T}}\; & \;\frac{1}{E_{L}}\;         & \;0\;                & \;0\; & \;0 \\ 
          \;0\;                     & \;0\;                       & \;0\;                       & \;\frac{1}{G_{RT}}\; & \;0\; & \;0 \\
          \;0\;                     & \;0\;                       & \;0\;                       & \;0\; & \;\frac{1}{G_{RL}}\; & \;0 \\
          \;0\;                     & \;0\;                       & \;0\;                       & \;0\; & \;0\; & \;\frac{1}{G_{TL}} \\     \end{array} \right].
\end{equation}
The non-zero off-diagonal terms of the elastic compliance matrix are mutually equal which is referred to as reciprocal dependencies and can be written by the following relations:
\begin{equation}
\frac{\nu_{RT}}{E_{R}}=\frac{\nu_{TR}}{E_{T}},~~~~~~\frac{\nu_{RL}}{E_{R}}=\frac{\nu_{LR}}{E_{L}},~~~~~~\frac{\nu_{TL}}{E_{T}}=\frac{\nu_{LT}}{E_{L}}.
\end{equation} 
Note that all engineering constants depend on the moisture level. They can be fitted by linear and third degree polynomial functions of moisture content~$\omega$ for the properties~$P_{b}$ and~$P_{s}$ associated with the species European beech and Norway spruce, respectively via parameters~$b_{x},s_{y}$ ($x$ = 0,1 and $y$ = 0,...,3): $P_b=b_0+b_1\omega$ and~$P_s=s_0+s_1\omega+s_2\omega^{2}+s_3\omega^{3}$ valid from oven-dry condition to the fiber saturated state. The relevant coefficients for describing material constants utilized in the definition of the compliance tensors are summarized in the \ref{Sec:Tables} in Table~\ref{table:Elascoef}.
\nomenclature{$P_{b}/P_{s}$}{European beech/Norway spruce property}
\subsubsection{Plastic deformation}\label{SSSec:Plastic}
Wood is a hygroscopic material with strong dependence of stiffness and strength on the moisture content. It is therefore prone to accumulate irrecoverable deformations even under combinations of moderate load with simultaneous moisture increase. The tendency of important hardwood species, such as European beech for high moisture sorption amplifies the drop of mechanical properties - strength in particular - what can result in excessive plastic deformations. Hence for reliable and meaningful predictions of the behavior of wood in constructions, it is advisable to consider irrecoverable constituents of the total strain.

In the last decade, significant efforts were made for describing the elasto-plastic mechanical behavior of wood based on the progress in metal plasticity. Experiments were conducted for uni- and bi-axial loading under tension and compression \cite{Eberhard02,Fleischmann03,Fleischmann05,HelnEberMan05} for different species and moisture content that lead to the following striking observations:
\begin{itemize}
	\item[(a)] Failure of wood under tensile or shear loading exhibits localized brittle fracture, however under compression, pronounced inelastic behavior is witnessed.
	\item[(b)] Under compression, two consecutive regimes are observed \cite{HelnMultisurf03,SchKaliMulti06}, namely cellular collapse, and for larger inelastic strains densification or compaction of the collapsed cells \cite{Fleischmann03,Fleischmann05}. 
	\item[(c)] Plastic hardening in different anatomical directions is only weakly coupled, since different orientation dependent micro-mechanical mechanisms act on the cellular scale.
\end{itemize}
These experimental observations have consequences for elasto-plastic models, the shape and type of yield surfaces and their evolution. Simple models use a single deformable 3D ellipsoidal yield surface \cite{HanHelnII2003,HelnShell05}, however despite of the advantage of having a closed surface with C2-consistency, observations (a,c) point at the limited validity of such an approach. The other extreme is given by a multi-surface approach, where seven distinct failure mechanisms and subsequently seven yield flow criteria comprising three tensile and three compressive failures along orthotropy directions together with one shear failure are considered \cite{SchKaliMulti06}. The constructed surface is only C0-consistent, requiring procedures for stress states that lie on intersections, since the evolution of the individual surfaces is not coupled, leading to 7 internal state variables. The model is basically a generalization of the presented plane stress model \cite{HelnMultisurf03} to 3D. In principle the brittle tensile response can be treated in the framework of plasticity as a smeared crack, however other approaches e.g., using cohesive elements are more stable \cite{SchmiKalisk09,ReschKalis10}. In later versions, therefore only the 3 compressive surfaces were used by the authors \cite{SchmiKalisk09}. A further reduction was proposed by \cite{ReschKalis10,SaftKalis11,HeringSaft12} with smooth corners (C1-continuity) and a combined evolution of all surfaces being described by a single strain-type internal state variable.

This study uses a yield surface similar to Ref.~\cite{SchmiKalisk09}, namely a three-dimensional orthotropic non-smooth multi-surface plasticity model (C0-continuity) consisting of three independent failure mechanisms along anatomical directions for compressive loading. To include the role of changing moisture content on the development of plasticity, all relevant strength values and respective hardening parameters are defined to be moisture dependent. The transition between elastic and inelastic domains in the stress space is characterized by three yield functions in the same form as the second-order polynomial failure criterion proposed by Ref.~\cite{TsaiWu71} as follows:
\begin{equation}     \label{eq:YFun}
f_{l}(\boldsymbol{\sigma},\alpha_{l},\omega)=\boldsymbol{\mathrm{a}}_{l}(\omega):\boldsymbol{\sigma}+\boldsymbol{\sigma}:\boldsymbol{\mathrm{b}}_{l}(\omega):\boldsymbol{\sigma}+q_{l}(\alpha_{l},\omega)-1,~~~~~~~~l=\text{R, T, L}.
\end{equation}
$\alpha_{l}$ denotes the strain-type internal state variable related to every anatomical direction,~$\boldsymbol{\mathrm{a}}_{l}$ and~$\boldsymbol{\mathrm{b}}_{l}$ resemble the strength tensors to be defined in the following, and eventually~$q_{l}$ is a scalar value for plastic hardening. 
 
Quantitative experimental data for the moisture dependence of the plastic hardening behavior is rather sparse. To characterize the hardening behavior we therefore adopt a mathematical approach proposed by Ref.~\cite{HeringSaft12} with a modified form of Ramberg-Osgood equations \cite{SimoHughes98} that was successfully applied to the compressive behavior of European beech at various moisture contents. \emph{Unidirectional} moisture-dependent isotropic hardening laws are applied that describe measured constitutive behavior under uni-axial compression along three anatomical orientations. The term ''\emph{unidirectional}'' emphasizes the independence of the hardening phenomenon of different failure mechanisms from each other, described above. The moisture-dependent hardening responses derived from the modified Ramberg-Osgood curves (see Ref.~\cite{HeringSaft12}) are approximated by exponential functions as follows:
\begin{equation}     \label{eq:Harden}
q_{l}(\alpha_{l},\omega)=(\beta_{0l}\omega+\beta_{1l})(1-e^{-\beta_{2l}\alpha_{l}})/f_{c,l}(\omega),
\end{equation}
in accordance with the available experimental results. Here~$l$ denotes the orientation R, T, L,~$\beta_{0l}$,~$\beta_{1l}$, and~$\beta_{2l}$ are material constants, and~$f_{c,l}(\omega)$ is the compressive strength of the material at the current level of moisture content. Since the last term of the right-hand-side of Eq.~(\ref{eq:YFun}) is equal to one, the current value of the hardening, i.e.,~$q_{l}(\alpha_{l},\omega)$ must be normalized by the compressive strengths to make it consistent with the dimension of the yield criterion expression. All parameters needed for the calculation of hardening functions for spruce and beech are summarized in the \ref{Sec:Tables} in Table~\ref{table:Hardparam}. Note that in this work contrary to Refs.~\cite{HelnMultisurf03,SchKaliMulti06} the densification regime is omitted (observation(b)).
\nomenclature{$f_{l}$}{Yield function,~$l=$R, T, L}
\nomenclature{$\boldsymbol{\mathrm{a}}_{l}, \boldsymbol{\mathrm{b}}_{l}$}{Strength tensors}
\nomenclature{$q_{l}$}{Plastic hardening function,~$l=$R, T, L}
\nomenclature{$f_{c,l}$}{Normal compressive strength of the material,~$l=$R, T, L}
\nomenclature{$f_{t,L}$}{Normal longitudinal tensile strength}
\nomenclature{$f_{s,RT/RL/TL}$}{Shear strength in RT/RL/TL plane}

Strength values are described by a linear dependence on the moisture content~$\omega$. Analogous to Table~\ref{table:Elascoef}, all corresponding properties for beech and spruce are calculated as~$P_b=(z_{b0}\omega+z_{b1})$ and~$P_s=(z_{s0}\omega+z_{s1})$, respectively where~$z_{b0}$,~$z_{b1}$,~$z_{s0}$, and~$z_{s1}$ are given in Table~\ref{table:Strenparam} in the \ref{Sec:Tables}. The first index in the symbolic presentations of the strength values ($c$,~$t$,~$s$) implies compressive, tensile, and shear, while the second one indicates either one of the anatomical directions or the corresponding plane. In the following, all strength tensors, i.e.,~$\boldsymbol{\mathrm{a}}_{l}$ and~$\boldsymbol{\mathrm{b}}_{l}$ required for the formulation of the yield criterion belonging to each failure mechanism based on the approach introduced in Ref.~\cite{SchKaliMulti06} and in accordance with the RTL alignment of the orthotropic material coordinate system are given.~$\boldsymbol{\mathrm{b}}_{l}$ are diagonal matrices with entries outside the main diagonal being zero. Note that all compressive yield stresses in the definition of the strength tensors are accompanied by a minus sign.
\begin{subequations}
\paragraph{Compression in radial direction}
\renewcommand{\arraystretch}{1.5}
\begin{align}
     \nonumber
     \boldsymbol{\mathrm{a}}_{R}&=\left[ \begin{array}{cccccc}
          \frac{1}{f_{c,R}}\;& \;0\;& \;0\; & \;0\; & \;0\; & \;0 \\ \end{array} \right]^\mathrm{T},  ~~~~~\text{for European beech and Norway spruce},     \\
\boldsymbol{\mathrm{b}}_{R}&= diag \left[ \begin{array}{cccccc}
					\;0, & \;\frac{0.0805}{f_{c,T}^{2}},\; & \;\frac{-0.1490}{f_{c,L}f_{t,L}},\; & \;\frac{0.1080}{f_{s,RT}^{2}},\; & \;\frac{0.1125}{f_{s,RL}^{2}},\; & \;\frac{0.0705}{f_{s,TL}^{2}}\; \\     \end{array} \right],~~~~~\text{for European beech},     \label{eq:aR1}\\
\boldsymbol{\mathrm{b}}_{R}&= diag \left[ \begin{array}{cccccc}
					\;0, & \;\frac{0.4000}{f_{c,T}^{2}},\; & \;\frac{-0.2500}{f_{c,L}f_{t,L}},\; & \;\frac{0.4000}{f_{s,RT}^{2}},\; & \;\frac{0.3300}{f_{s,RL}^{2}},\; & \;\frac{0.3300}{f_{s,TL}^{2}}\; \\     \end{array} \right],~~~~~\text{for Norway spruce}.
\end{align}
\end{subequations}
\begin{subequations}
\paragraph{Compression in tangential direction}
\renewcommand{\arraystretch}{1.5}
\begin{align}
     \nonumber
     \boldsymbol{\mathrm{a}}_{T}&=\left[ \begin{array}{cccccc}
          \;0\; & \frac{1}{f_{c,T}}\; & \;0\; & \;0\; & \;0\; & \;0 \\ \end{array} \right]^\mathrm{T},  ~~~~~\text{for European beech and Norway spruce},      \\
\boldsymbol{\mathrm{b}}_{T}&= diag \left[ \begin{array}{cccccc}
          \;\frac{0.0805}{f_{c,R}^{2}},\; & \;0,\; & \;\frac{-0.1490}{f_{c,L}f_{t,L}},\; & \;\frac{0.1080}{f_{s,RT}^{2}},\; & \;\frac{0.1125}{f_{s,RL}^{2}},\; & \;\frac{0.0705}{f_{s,TL}^{2}}\; \\     \end{array} \right],~~~~~\text{for European beech},     \label{eq:aT1}\\
\boldsymbol{\mathrm{b}}_{T}&= diag \left[ \begin{array}{cccccc}
          \;\frac{0.4000}{f_{c,R}^{2}},\; & \;0,\; & \;\frac{-0.2500}{f_{c,L}f_{t,L}},\; & \;\frac{0.4000}{f_{s,RT}^{2}},\; & \;\frac{0.3300}{f_{s,RL}^{2}},\; & \;\frac{0.3300}{f_{s,TL}^{2}}\; \\     \end{array} \right],~~~~~\text{for Norway spruce}.          
\end{align}
\end{subequations}
\begin{subequations}
\paragraph{Compression in longitudinal direction}
\renewcommand{\arraystretch}{1.5}
\begin{align}
     \nonumber
     \boldsymbol{\mathrm{a}}_{L}&=\left[ \begin{array}{cccccc}
          \;0\; & \;0\; & \frac{1}{f_{c,L}}\; & \;0\; & \;0\; & \;0 \\ \end{array} \right]^\mathrm{T},   ~~~~~\text{for European beech and Norway spruce},        \\
\boldsymbol{\mathrm{b}}_{L}&= diag \left[ \begin{array}{cccccc}
          \;\frac{0.0665}{f_{c,R}^{2}},\; &  \;\frac{0.0665}{f_{c,T}^{2}},\;  & \;0,\; & \;\frac{0.0675}{f_{s,RT}^{2}},\; & \;\frac{0.0855}{f_{s,RL}^{2}},\; & \;\frac{0.0530}{f_{s,TL}^{2}}\; \\     \end{array} \right],~~~~~\text{for European beech},          \label{eq:aL1}\\
\boldsymbol{\mathrm{b}}_{L}&= diag \left[ \begin{array}{cccccc}
          \;\frac{0.3300}{f_{c,R}^{2}},\; &  \;\frac{0.3300}{f_{c,T}^{2}},\;  & \;0,\; & \;\frac{0.2500}{f_{s,RT}^{2}},\; & \;\frac{0.2500}{f_{s,RL}^{2}},\; & \;\frac{0.2500}{f_{s,TL}^{2}}\; \\     \end{array} \right],~~~~~\text{for Norway spruce}.          
\end{align}
\end{subequations}

\begin{figure*}[htpb]
     \begin{center}
\includegraphics[width=1\textwidth]{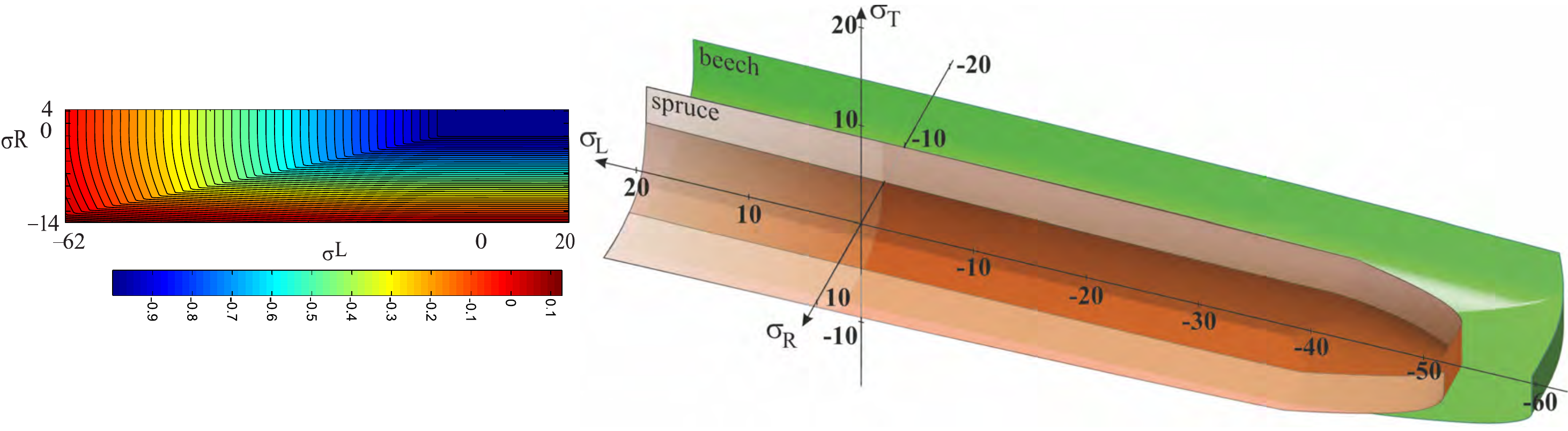} 
    \caption{Left: 2D illustration of the yield surfaces in the R-L plane and for~$\sigma_{RL}=0$. Right: 3D visualization of the boundaries between elastic and plastic domains in the principal stress space. Both are illustrated for~$q_{l}=0$ at~$\omega=12\%$.}
   \label{fig:YSurfStrss}
   \end{center}
\end{figure*}

The scalar coefficients in the numerator of the diagonal components of the strength matrices~$\boldsymbol{\mathrm{b}}_{l}$ are weighting factors which following Ref.~\cite{SchKaliMulti06} can be adjusted to the respective species. To make an adaptation from spruce (s) to beech (b), we replace the respective strength value of the denominator by~$f^s=f^b\cdot\left(f^b_{12}/f^s_{12}\right)$ what corresponds to a scaling of the scalar values with strength value ratios at~$\omega=12\%$. Until today, bi-axial tests for beech have not been published in literature, hence a final justification for this adaptation assumption is not possible. In Fig.~\ref{fig:YSurfStrss} the boundary between the linear elastic domain and the non-linear behavior of the material, based on strength tensors~$\boldsymbol{\mathrm{a}}_{l}$ and~$\boldsymbol{\mathrm{b}}_{l}$ at~$\omega=12\%$, and for~$\left(q_{l}=0\right)$ is shown. Because of the three-dimensional representation of the stress tensor, the demonstration of all yield surfaces for any arbitrary stress state through one individual image is not feasible. Accordingly, in Fig.~\ref{fig:YSurfStrss} (left) a two-dimensional illustration of the yield conditions under planar state of stress in the R-L plane, i.e.,~$(\sigma_{T}=\sigma_{RT}=\sigma_{TL}=0)$ and for principal normal stresses~$\left(\sigma_{RL}=0\right)$ is depicted. Fig.~\ref{fig:YSurfStrss} (right) shows a three-dimensional visualization of the failure surfaces for the situation in which principal directions of stress and axes of the local material coordinate system are coincident~$(\sigma_{RT}=\sigma_{RL}=\sigma_{TL}=0)$.

By now, all moisture-dependent parameters needed for the description of the rate-independent multi-surface plasticity material model are introduced. The general algorithm for the numerical implementation of the above-mentioned model is explained in detail in Refs.~\cite{SimoHughes98,BazantJirasek02}. Now we focus on a concise description of the principles of the evolution equations for irrecoverable deformations in the context of non-smooth multi-surface plasticity. The constitutive relation of the plastic deformation under the assumption of a standard associative plasticity is described by the following formula known as flow rule, usually referred to as \emph{Koiter's} rule \cite{Koiter53}:
\begin{equation}     \label{eq:Plasflow}
\dot{\boldsymbol{\varepsilon}}^{pl} =\sum_{l=1}^{r}\dot{\gamma}^{l} \partial_{\boldsymbol{\sigma}}f_{l}(\boldsymbol{\sigma},\alpha_{l},\omega)=\sum_{l=1}^{r}\dot{\gamma}^{l}(\boldsymbol{\mathrm{a}}_{l}(\omega)+2\boldsymbol{\mathrm{b}}_{l}(\omega):\boldsymbol{\sigma}).
\end{equation}
\nomenclature{$r/r_{adm}$}{Number of active yield mechanisms/Number of active yield constraints}
\nomenclature{$\partial_{\boldsymbol{\sigma}}f_{l}$}{Plastic flow direction tensor,~$l=$R, T, L}
\nomenclature{$H_{l}$}{Hardening modulus,~$l=$R, T, L}
\nomenclature{$\dot{\gamma}^{l}$}{Rate form of the plastic consistency parameter,~$l=$R, T, L}
\nomenclature{$\mathrm{\mathbb{S}}_{adm/act}$}{Set of admissible/active yield constraints}
Here~$r$ denotes the number of active yield mechanisms. Similarly with associative hardening, the evolution equation corresponding to the hardening law takes the form
\begin{equation}     \label{eq:Hardlaw}
\dot{\boldsymbol{\alpha}} = \sum_{l=1}^{r}\dot{\gamma}^{l} \partial_{\boldsymbol{q}}f_{l}(\boldsymbol{\sigma},\alpha_{l},\omega) =\sum_{l=1}^{r}\dot{\gamma}^{l}.
\end{equation}
In Eqs.~(\ref{eq:Plasflow}) and~(\ref{eq:Hardlaw}),~$\dot{\gamma}^{l}$ are plastic consistency parameters, which fulfill the following \emph{Kuhn-Tucker} complementary requirements:
\begin{equation}     \label{eq:Kuhn}
\dot{\gamma}^{l}\geq{0},~~~~~f_{l}(\boldsymbol{\sigma},\alpha_{l},\omega)\leq{0},~~~~~\dot{\gamma}^{l}f_{l}(\boldsymbol{\sigma},\alpha_{l},\omega)=0,     \\
\end{equation}
together with the consistency condition:
\begin{equation}     \label{eq:Cons}
\dot{\gamma}^{l}\dot{f_{l}}(\boldsymbol{\sigma},\alpha_{l},\omega)=0.     \\
\end{equation}
For an assumed number of active yield constraints~$r_{adm}$ and a given state of stress and internal hardening variable(s)
\begin{equation}     \label{eq:Admis}
\mathrm{\mathbb{S}}_{adm}:=\left\{l~\in~\left\{R,T,L\right\}~|~f_{l}(\boldsymbol{\sigma},\alpha_{l},\omega)=0\right\},     \\
\end{equation}
the \emph{Kuhn-Tucker} complementary conditions also known as loading/unloading requirements can be restated for multi-surface plasticity as the following \cite{SimoHughes98}:
\begin{description}
	\item[Condition 1.] If~$f_{l}(\boldsymbol{\sigma},\alpha_{l},\omega)<0$ or~$f_{l}(\boldsymbol{\sigma},\alpha_{l},\omega)=0$ and~$\dot{f_{l}}(\boldsymbol{\sigma},\alpha_{l},\omega)<0$, subsequently~$\dot{\gamma}^{l}=0$, which define the case of elastic loading or unloading, where plastic strains along with hardening variable(s) do not change.  
	\item[Condition 2.] If~$f_{l}(\boldsymbol{\sigma},\alpha_{l},\omega)=0$ and~$\dot{f_{l}}(\boldsymbol{\sigma},\alpha_{l},\omega)=0$, then~$\dot{\gamma}^{l}\geq0$, which specifies the plastic loading and subsequently the evolution of plastic deformation and respective internal variable(s).
\end{description}
Accordingly, based on Eq.~(\ref{eq:Admis}) and an expanded form of the loading/unloading conditions,~$1 \le r \le r_{adm}$ representing the number of active yield conditions reads as:
\begin{equation}     \label{eq:Activ}
\mathrm{\mathbb{S}}_{act}:=\left\{l~\in~\mathrm{\mathbb{S}}_{adm}~|~\dot{f_{l}}(\boldsymbol{\sigma},\alpha_{l},\omega)=0\right\}.     \\
\end{equation}

To summarize, flow role Eq.~(\ref{eq:Plasflow}) and hardening law Eq.~(\ref{eq:Hardlaw}) in combination with \emph{Kuhn-Tucker} complementary requirements form a set of non-linear equations with~$\dot{\gamma}^{l}$ as unknown variables. The corresponding solution can be realized through an iterative numerical procedure like a \emph{Newton-Raphson} scheme \cite{HanHelnI2003,HanHelnII2003,SimoHughes98}.   
\subsubsection{Hygro-expansion}\label{SSSec:Hygro}
Hygro-expansion describes swelling or shrinkage of the material under varying moisture content. Analogous to thermal expansion with respect to temperature gradients, it is assumed to be proportional to increments of moisture content with
\begin{equation}
\boldsymbol{\varepsilon}^{\omega}=\boldsymbol{\alpha}_{\omega}(Min(\omega,\omega_{FS})-\omega_{\textit{0}}),
\end{equation}
with the current moisture content~${\omega}$ and the fiber saturation moisture level~${\omega_{FS}}$ above which no hygro-expansion occurs.~$\omega_{\textit{0}}$ signifies the initial reference moisture content. The vector~$\boldsymbol{\alpha}_{\omega}$ contains hygro-expansion coefficients along orthotropy directions in an RTL material coordinate system defined by~$\boldsymbol{\alpha}_{\omega}=\left\{\alpha_{R}, \alpha_{T}, \alpha_{L}, 0, 0, 0\right\}^\mathrm{T}$. We assume that hygro-expansion coefficients are constant and independent from moisture variations. Corresponding values for European beech \cite{HeringThes11} and spruce \cite{GerekeThes09,NeuhausThes81} are given in the \ref{Sec:Tables} in Table~\ref{table:Hygcoef}. 
\nomenclature{$\omega/\omega_{FS}/\omega_{\textit{0}}$}{Current moisture content/Fiber saturation/Initial reference moisture level in \%}
\nomenclature{$\boldsymbol{\alpha}_{\omega}$}{Hygro-expansion coefficients tensor}
\subsubsection{Visco-elastic creep}\label{SSSec:Visco} 
Wood with constant moisture content and under sustained loading exhibits time-dependent deformation generally termed visco-elastic creep \cite{HanHelnI2003,HanHelnII2003,Fortino09}. We describe visco-elastic behavior based on a fully recoverable approach by serial association of \emph{Kelvin-Voigt} elements. It's noteworthy to mention that the following formulations are given within the framework of linear visco-elasticity and are valid for the first stage of the feature known as primary step. 

By taking the derivative of the free energy function Eq.~(\ref{eq:Psi}) with respect to the element-wise visco-elastic strain, the driving stress for the~$i^{th}$ visco-elastic \emph{Kelvin-Voigt} element is obtained as
\begin{equation}     \label{eq:Viselas}
\boldsymbol{\sigma}_{i}^{ve}=-\frac{\partial\psi}{\partial\boldsymbol{\varepsilon}_{i}^{ve}}={\bf C}_{0}:\left(\boldsymbol{\varepsilon}-\boldsymbol{\varepsilon}^{pl}-\boldsymbol{\varepsilon}^{\omega}-\sum_{i=1}^{n} \boldsymbol{\varepsilon}_{i}^{ve}-\sum_{j=1}^{m} \boldsymbol{\varepsilon}_{j}^{ms}\right)-{\bf C}_{i}:\boldsymbol{\varepsilon}_{i}^{ve}=\boldsymbol{\sigma}-{\bf C}_{i}:\boldsymbol{\varepsilon}_{i}^{ve}.
\end{equation}
In this relation~${\bf C}_{i}$ stands for the visco-elastic stiffness matrix. The visco-elastic strain rate~$\dot{\boldsymbol{\varepsilon}}_{i}^{ve}$ is considered to be a linear function of visco-elastic driving stress and reads:
\begin{equation}
     \label{eq:Vrate}
\dot{\boldsymbol{\varepsilon}}_{i}^{ve}=\frac{1}{\tau_{i}}{\bf C}_{i}^{-1}:\boldsymbol{\sigma}_{i}^{ve}.
\end{equation}
Subsequently, the governing rate equation for an individual visco-elastic \emph{Kelvin-Voigt} element is
\begin{equation}
     \label{eq:Visrate}
\dot{\boldsymbol{\varepsilon}}_{i}^{ve}+\frac{1}{\tau_{i}}\boldsymbol{\varepsilon}_{i}^{ve}=\frac{1}{\tau_{i}}{\bf C}_{i}^{-1}:\boldsymbol{\sigma}(t),
\end{equation}
where~${\bf C}_{i}^{-1}$ and~$\tau_{i}$ denote the visco-elastic compliance tensor and the characteristic retardation time relevant to the~$i^{th}$ \emph{Kelvin-Voigt} element, respectively. Following Refs.~\cite{HanHelnI2003,HanHelnII2003,Fortino09}, the visco-elastic compliance tensor is assumed to be proportional to the elastic compliance matrix with a unitless scalar~$\gamma_{i}^{ve}$, namely  
\begin{equation}
     \label{eq:Gammai}
\gamma_{i}^{ve}={\bf C}_{0}^{-1}/{\bf C}_{i}^{-1}~.
\end{equation}
For spruce, the dimensionless fractions Eq.~(\ref{eq:Gammai}) are taken from Ref.~\cite{Fortino09}, while for beech they are calculated based on creep measurements of component of the visco-elastic compliance tensor at different moisture levels in grain~$J_{L}^{ve}$ \cite{HeringCreep12} with linear moisture dependence. For a serial combination of four \emph{Kelvin-Voigt} elements, Table~\ref{table:Gamval} in the \ref{Sec:Tables} provides the parameters that describe the moisture dependent longitudinal component of the creep compliance tensor for the~$i^{th}$ \emph{Kelvin-Voigt} element:
\begin{equation}
     \label{eq:LCompl}
J_{iL}^{ve}=(J_{i1}\omega+J_{i0})(1-e^{-t/\tau_{i}}).
\end{equation}
The ratio of the longitudinal component of the elastic compliance tensor~$\frac{1}{E_{L}}$ with this value gives the fraction~$\gamma_{i}^{ve}$ relevant to each \emph{Kelvin-Voigt} element. Note that due to sparse experimental data, it is a common practice to apply the fraction~$\gamma_{i}^{ve}$ measured in the grain also to the cross-grain directions. Additionally, the same value of the characteristic time (viscosity) along the grain is utilized for other anatomical directions as well. Hence retardation times are defined isotropically.
 
Following Refs.~\cite{HanHelnI2003,HanHelnII2003,Fortino09} by integrating Eq.~(\ref{eq:Visrate}), the element-wise visco-elastic strain response of the~$i^{th}$ \emph{Kelvin-Voigt} element reads 
\begin{equation}
     \label{eq:VElasSol}
\boldsymbol{\varepsilon}_{i,n+1}^{ve}=\boldsymbol{\varepsilon}_{i,n}^{ve}~\mathrm{exp}(-\frac{\Delta{t}}{\tau_{i}})+\int_{t_{n}}^{t_{n+1}} \frac{{\bf C}_{i}^{-1}:\boldsymbol{\sigma}(t)}{\tau_{i}}~\mathrm{exp}(-\frac{t_{n+1}-t}{\tau_{i}})~\,\mathrm{d}t, 
\end{equation}
for a stress driven problem, with the time step~$\Delta{t}=t_{n+1}-t_{n}$ and the visco-elastic strain tensors~$\boldsymbol{\varepsilon}_{i,n+1}^{ve},~\boldsymbol{\varepsilon}_{i,n}^{ve}$ at time~$t_{n+1}$ and~$t_{n}$, respectively.
\subsubsection{Mechano-sorptive creep}\label{SSSec:Mechano}
\begin{figure*}[htpb]
	  \begin{center}
	\includegraphics[scale=0.75]{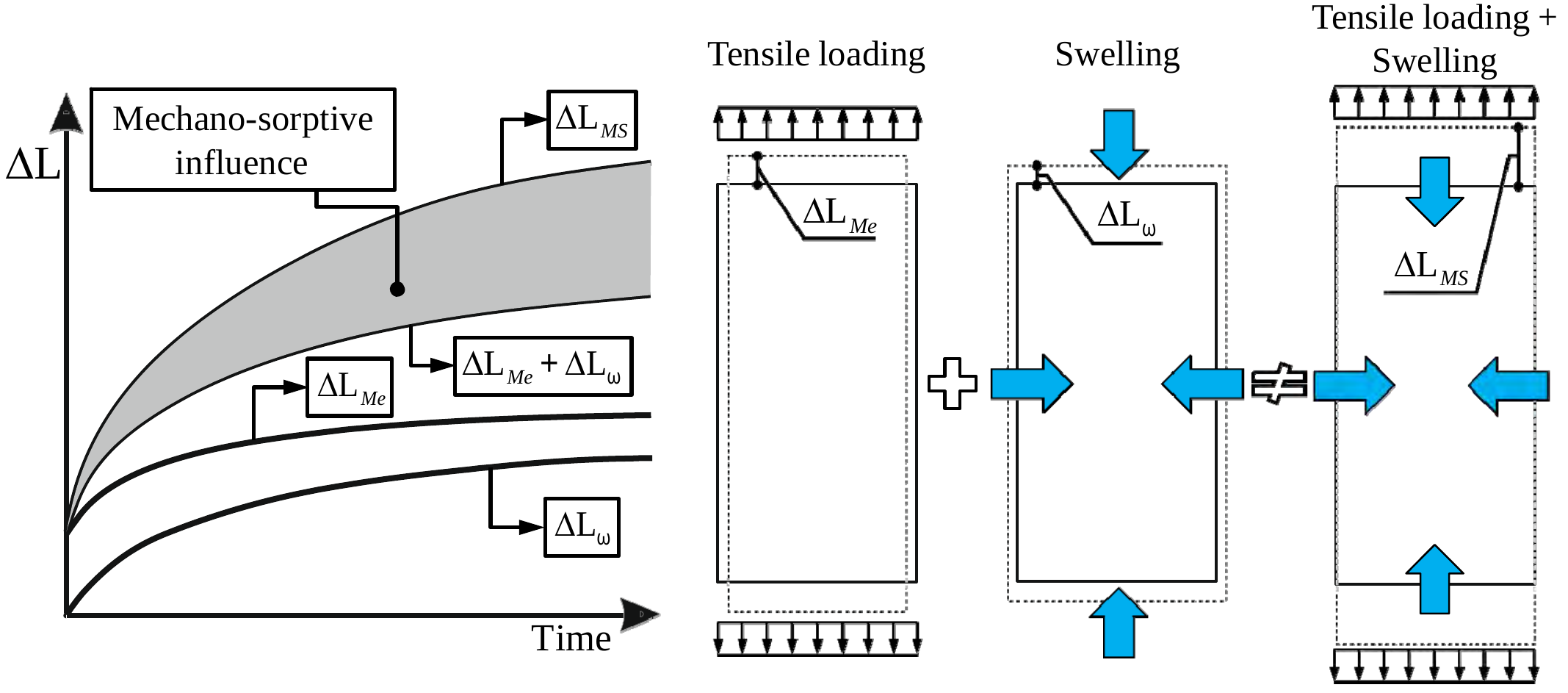}
	\caption{Schematic interpretation of the mechano-sorptive creep.~$\Delta\mathrm{L}_{Me}$,~$\Delta\mathrm{L}_{\omega}$,and~$\Delta\mathrm{L}_{MS}$ denote increase in the length caused by external mechanical loading, swelling, and mechano-sorptive effect, respectively.} 
	\label{fig:MechConcept}
	  \end{center}
\end{figure*}

The effect that the observed deformation of a loaded specimen under changing moisture content is remarkably higher than the deformation of a loaded specimen under constant climatic conditions superimposed by the deformation of an unloaded specimen under varying moisture level is known as mechano-sorptive creep (see Fig.~\ref{fig:MechConcept}). Experimental studies for spruce can be found in Refs.~\cite{RantaMau75,Martensson94,HanhiMech00a}, while for beech such studies are still missing. 

For the numerical description of mechano-sorption in principle \emph{Kelvin-Voigt} type elements are used \cite{HanHelnI2003,HanHelnII2003,Fortino09}. In Refs.~\cite{HanHelnI2003,HanHelnII2003} plasticity is part of mechano-sorption, while in the formulation by Ref.~\cite{Fortino09} that we follow, it is an independent strain contribution. After differentiating the free energy function Eq.~(\ref{eq:Psi}) with respect to the element-wise mechano-sorptive creep strain, the corresponding driving stress becomes 
\begin{equation}
     \label{eq:MechSorp}
\boldsymbol{\sigma}_{j}^{ms}=-\frac{\partial\psi}{\partial\boldsymbol{\varepsilon}_{j}^{ms}}={\bf C}_{0}:\left(\boldsymbol{\varepsilon}-\boldsymbol{\varepsilon}^{pl}-\boldsymbol{\varepsilon}^{\omega}-\sum_{i=1}^{n} \boldsymbol{\varepsilon}_{i}^{ve}-\sum_{j=1}^{m} \boldsymbol{\varepsilon}_{j}^{ms}\right)-{\bf C}_{j}:\boldsymbol{\varepsilon}_{j}^{ms} =\boldsymbol{\sigma}-{\bf C}_{j}:\boldsymbol{\varepsilon}_{j}^{ms},
\end{equation}
where~${\bf C}_{j}$ stands for the mechano-sorptive stiffness tensor. For characterizing mechano-sorption, a rate equation quite similar to the one of the visco-elastic creep Eq.~(\ref{eq:Vrate}) is applied:
\begin{equation}
     \label{eq:Mrate}
\dot{\boldsymbol{\varepsilon}}_{j}^{ms}=\frac{\left|\dot{\omega}\right|}{\mu_{j}}{\bf C}_{j}^{-1}:\boldsymbol{\sigma}_{j}^{ms}.
\end{equation}
Here~$\mu_{j}$ is called characteristic moisture analogous to the characteristic retardation time in visco-elasticity and~${\bf C}_{j}^{-1}$ designates the tensor of mechano-sorptive compliance respective to the~$j^{th}$ \emph{Kelvin-Voigt} element. By inserting Eq.~(\ref{eq:MechSorp}) in Eq.~(\ref{eq:Mrate}) and rearranging, the constitutive relation for a single mechano-sorptive \emph{Kelvin-Voigt} type element can be written as
\begin{equation}
     \label{eq:Mechrate}
\dot{\boldsymbol{\varepsilon}}_{j}^{ms}+\frac{\left|\dot{\omega}\right|}{\mu_{j}}\boldsymbol{\varepsilon}_{j}^{ms}=\frac{\left|\dot{\omega}\right|}{\mu_{j}}{\bf C}_{j}^{-1}:\boldsymbol{\sigma}(t).
\end{equation}
Note that the solution of Eq.~(\ref{eq:Mechrate}) is identical to visco-elasticity Eq.~(\ref{eq:Visrate}), but instead of a time increment the value of the absolute moisture content change is taken. Consequently, like in Eq.~(\ref{eq:VElasSol}), the mechano-sorptive strain is
\begin{equation}
     \label{eq:MechSol}
\boldsymbol{\varepsilon}_{j,n+1}^{ms}=\boldsymbol{\varepsilon}_{j,n}^{ms}~\mathrm{exp}(-\frac{\left|\Delta{\omega}\right|}{\mu_{j}})+\int_{\omega_{n}}^{\omega_{n+1}} \frac{{\bf C}_{j}^{-1}:\boldsymbol{\sigma}\left(t\right)}{\mu_{j}}~\mathrm{exp}(-\frac{\left|\omega_{n+1}-\omega(t)\right|}{\mu_{j}})~\,\mathrm{d}\left|\omega(t)\right|. 
\end{equation}
Here~$\left|\Delta{\omega}\right|=\left|\omega_{n+1}-\omega_{n}\right|$ is the absolute value of moisture increment,~$\boldsymbol{\varepsilon}_{j,n+1}^{ms}$ and~$\boldsymbol{\varepsilon}_{j,n}^{ms}$ denote the mechano-sorptive strains evaluated at the times~$t_{n+1}$ and~$t_{n}$, respectively. In order to calculate the element-wise mechano-sorptive compliance, a similar approach as for the visco-elastic counterpart is utilized. The corresponding scalar fractions, i.e.,~$\gamma_{j}^{ms}$ are obtained based on the tangential and longitudinal components of the mechano-sorptive compliance tensor presented in Ref.~\cite{Fortino09} for spruce defined as
\begin{equation}
     \label{eq:Gammaj}
\gamma_{j}^{ms}={\bf C}_{0}^{-1}/{\bf C}_{j}^{-1}~.
\end{equation}
Following the work of~\cite{Fortino09}, who calibrated the values for three serial \emph{Kelvin-Voigt} elements on experimental data from Refs.~\cite{Toratti92,ToratSven02}, the element-wise mechano-sorptive compliance tensor takes the following form:  
\renewcommand{\arraystretch}{1.5}
\begin{equation}
     \label{eq:Mechcomp}
     {\bf C}_{j}^{-1}=\left[ \begin{array}{cccccc}
          \frac{J_{jT0}^{ms}E_{T0}}{E_{R}}\;& \;-\frac{J_{jT0}^{ms}E_{T0}\nu_{TR}}{E_{T}}\;& \;-\frac{J_{jT0}^{ms}E_{T0}\nu_{LR}}{E_{L}}\; & \;0\; & \;0\; & \;0 \\
         -\frac{J_{jT0}^{ms}E_{T0}\nu_{RT}}{E_{R}}\; & \;\frac{J_{jT0}^{ms}E_{T0}}{E_{T}}\; & \;-\frac{J_{jT0}^{ms}E_{T0}\nu_{LT}}{E_{L}}\; & \;0\; & \;0\; & \;0 \\
         -\frac{J_{jT0}^{ms}E_{T0}\nu_{RL}}{E_{R}}\; & \;-\frac{J_{jT0}^{ms}E_{T0}\nu_{TL}}{E_{T}}\; & \;J_{jL}^{ms}\; & \;0\; & \;0\; & \;0 \\ 
          \;0\;                     & \;0\;                       & \;0\;     & \;\frac{J_{jT0}^{ms}E_{T0}}{G_{RT}}\; & \;0\; & \;0 \\
          \;0\;                     & \;0\;                       & \;0\;     & \;0\; & \;\frac{J_{jT0}^{ms}E_{T0}}{G_{RL}}\; & \;0 \\
          \;0\;                     & \;0\;                       & \;0\;     & \;0\; & \;0\; & \;\frac{J_{jT0}^{ms}E_{T0}}{G_{TL}} \\     \end{array} \right].
\end{equation}
The values corresponding to~$J_{jT0}^{ms}$ and~$J_{jL}^{ms}$ together with the characteristic moistures~$\mu_{j}$ are summarized in the \ref{Sec:Tables} in Table~\ref{table:Mechprops}. Due to lack of data on the mechano-sorptive behavior of beech, the corresponding values of the mechano-sorptive compliance tensor are calculated by scaling the spruce values by the ratio of densities of the two species~$\left(470/720\right)$~\cite{Niemz93}.
\nomenclature{$\boldsymbol{\sigma}/\boldsymbol{\sigma}_{i}^{ve}/\boldsymbol{\sigma}_{j}^{ms}$}{Total stress/$i^{th}$ visco-elastic stress/$j^{th}$ mechano-sorptive stress tensor}
\nomenclature{${\bf C}_{0}/{\bf C}_{i}/{\bf C}_{j}$}{Elastic stiffness/$i^{th}$ visco-elastic stiffness/$j^{th}$ mechano-sorptive stiffness tensor}
\nomenclature{${\bf C}_{0}^{-1}/{\bf C}_{i}^{-1}/{\bf C}_{j}^{-1}$}{Elastic compliance/$i^{th}$ visco-elastic compliance/$j^{th}$ mechano-sorptive compliance tensor}
\nomenclature{$\tau_{i}/\mu_{j}$}{$i^{th}$ characteristic retardation time/$j^{th}$ characteristic moisture}
\nomenclature{$\gamma_{i}^{ve}/\gamma_{j}^{ms}$}{$i^{th}$ scalar visco-elastic/$j^{th}$ scalar mechano-sorptive compliance fraction}
\nomenclature{$J_{iL}^{ve}$}{Longitudinal component of~$i^{th}$ visco-elastic compliance tensor}
\nomenclature{$\Delta{t}/t_{n+1},t_{n}$}{Time step/Time moments~$n$ and~$n+1$}
\nomenclature{$\left|\Delta{\omega}\right|$}{Absolute value of moisture increment}
\nomenclature{$J_{jL/jT0}^{ms}$}{Longitudinal/Tangential entry of the~$j^{th}$ mechano-sorptive compliance tensor}
\subsection{Moisture-stress analysis}\label{SSec:Moisture}
Due to the importance of moisture fields and gradients for the behavior of wood, moisture transport is essential for simulations. In this work mechanical and thermal fields do not influence moisture transport, so the problem is only partially coupled. Furthermore we assume moisture transport inside the porous material wood to be dominated by diffusive transport~\cite{GerekeThes09}. Experimental observations of moisture transport below fiber saturation show more correspondence with non-\emph{Fickian} behavior \cite{Wadso94,Krabbenhoft04}, or multi-\emph{Fickian} diffusion \cite{Frandsen07}. In the present work for the sake of simplicity and also because of insufficient details regarding either non-\emph{Fickian} or multi-\emph{Fickian} formulations, \emph{Fick}'s law for moisture transfer is employed. 

In accordance with the \emph{Fick}'s first law the body moisture flux is given by
\begin{equation}     \label{eq:FFick1}
{\bf J}_{\omega}^{b}=-{\bf D}\nabla{c},
\end{equation}
where~${\bf J}_{\omega}$ is body moisture flux vector,~{\bf D} denotes the matrix of diffusion coefficients,~$c=\rho_{0}\omega$ is the water concentration,~$\rho_{0}$ the oven-dry wood density, and~$\omega$ is again the moisture content in [\%]. \emph{Fick}'s second law expresses the change of the concentration with respect to the time. In a general form and for changing diffusion coefficients, the time variation of concentration is written as:
\begin{equation}     \label{eq:FFick2c}
\frac{\partial{c}}{\partial{t}}=\nabla.({\bf D}\nabla{c}).
\end{equation}
For constant density Eq.~(\ref{eq:FFick2c}) simplifies to
\begin{equation}     \label{eq:FFick2w}
\frac{\partial{\omega}}{\partial{t}}=\nabla.({\bf D}\nabla{\omega}).
\end{equation}
In general, mathematical formulations of moisture diffusion and heat transfer are similar. The \emph{Fourier} equations of heat transfer are
\begin{equation}        \label{eq:Fourier}
{\bf J}_{T}=-{\bf K}\nabla{T},  ~~~~~~~\text{and}~~~~~~~(\rho c_{T})\frac{\partial{T}}{\partial{t}}=\nabla.({\bf K}\nabla{T})~,
\end{equation}
with the heat flux~${\bf J}_{T}$, density~$\rho$, specific heat~$c_{T}$ and temperature~$T$, as well as the matrix of thermal conductivity coefficients~{\bf K}. The analogy between Eqs.~(\ref{eq:FFick2w})~and~(\ref{eq:Fourier}) is preserved when~$(\rho c_{T})=1$, rendering thermal analysis capabilities of FE packages valid for moisture transport simulations. In this work, the diffusion process is assumed to be uncoupled among anatomical directions and subsequently, the matrix of orthotropic diffusion coefficients is defined as:
\renewcommand{\arraystretch}{1.5}
\begin{equation}
     \label{eq:DCoeff}
     \mathrm{\bf{D}}=diag \left[ \begin{array}{ccc} \mathrm{D}_{R} & \mathrm{D}_{T} & \mathrm{D}_{L}  \end{array} \right].  
\end{equation}
The diffusion coefficients are considered to be moisture-dependent and can be described by exponential laws as a function of moisture, namely
\begin{equation}
     \label{eq:DoeffExp}
\mathrm{D}_{l}(\omega)=\mathrm{D}_{0l}~(e^{\alpha_{0l}\omega}),
\end{equation}
where~$l$=R, T, and L. All parameters required to calculate moisture-dependent diffusion coefficients for spruce and beech \cite{SaftKalis11,HeringThes11} are summarized in the \ref{Sec:Tables} in Table~\ref{table:Moisprops}. 
\nomenclature{${\bf J}_{\omega}/{\bf J}_{T}$}{Body moisture/Heat flux vector}
\nomenclature{$\text{{\bf D}}/\text{{\bf K}}$}{Tensor of diffusion coefficients/thermal conductivity coefficients}
\nomenclature{$\text{c}/\rho_{0}/\text{T}/c_{T}$}{Water concentration/Oven-dry wood density/Temperature/Specific heat}
\section{Numerical implementation of the moisture-dependent rheological model}\label{Sec:NumInt}
To implement the material model with the presented rheological behavior in the framework of a FE simulation, an incremental, iterative numerical approach is needed~\cite{HanHelnI2003,HanHelnII2003,Fortino09} with time increments~$(\Delta{t}=t_{n+1}-t_{n})$. In the following, subscripts~$(\bullet)_{n}$ indicate a state at the beginning of a time step, whereas the subscript~$(\bullet)_{n+1}$ refers to the end of the time increment. The stress update algorithm is based on the assumption that at time~$t=t_{n}$, the state of the material is available to be able to calculate the corresponding solution at time~$t=t_{n+1}$ by means of an incremental update procedure. In detail the state variables are the moisture distribution, the total strain including all five corresponding partial constituents, internal plastic hardening variables and the total stress tensor. The individual algorithmic tangent operators associated with each deformation mode are computed separately and then, by incorporating all single Jacobians, the tangent operator for the entire model is obtained following Refs.~\cite{HanHelnI2003,HanHelnII2003,Fortino09}. Additionally, the total strain increment and the amount of moisture content change are needed for the iteration. 
The hygro-expansion strain tensor~$\left(\boldsymbol{\varepsilon}_{n+1}^{\omega}\right)$ and the total strain tensor~$\left(\boldsymbol{\varepsilon}_{n+1}\right)$ at the end of the time increment are estimated and needed for the incrementation as well, using the old tangent operators. At each integration point the total stress tensor and total Jacobian matrix, meaning the algorithmic tangent operator for the whole constitutive model, are updated at the end of the time increment. In addition, the values of state variables in terms of elastic strain~$(\boldsymbol{\varepsilon}^{el})$, irrecoverable plastic deformation~$(\boldsymbol{\varepsilon}^{pl})$ along with related strain-type hardening variable(s)~$(\alpha_{l})$, viscoelastic strain tensor respective to every \emph{Kelvin-Voigt} element~$(\boldsymbol{\varepsilon}_{i}^{ve})$, and all element-wise mechano-sorptive strain tensors~$(\boldsymbol{\varepsilon}_{j}^{ms})$ have to be updated for the next increment. A brief overview of the iterative algorithm utilized for the decomposition of the total strain~$(\boldsymbol{\varepsilon})$ and the incremental procedure for the update of total stress and all state variables can be summarized as below:
\begin{enumerate}
  \item \textbf{Stage 1: Data initialization} Based on the iterative algorithm for the stress update, all state variables are set to their corresponding values from the last converged iteration of the previous increment as:
\begin{equation}
     \label{eq:VeMsInitval}
       \boldsymbol{\varepsilon}_{i,n+1}^{ve(k=0)}=\boldsymbol{\varepsilon}_{i,n}^{ve}, \qquad \boldsymbol{\varepsilon}_{j,n+1}^{ms(k=0)}=\boldsymbol{\varepsilon}_{j,n}^{ms}~.
\end{equation}
The superscript~$k$ refers to the~$\left(k\right)^{th}$ iteration of the considered increment. The first one ($k=0$) therefore is identical to the converged solution of the former increment.
  \item \textbf{Stage 2: Plastic deformation} In the next step, the possible development of irrecoverable deformation by plastic strain is examined. For this purpose a two-step return-mapping algorithm known as elastic predictor/plastic corrector based on the general multi-surface closest point projection approach is used \cite{SimoHughes98,BazantJirasek02}. According to the trial state of elastic strain at the end of the time increment~$\left(\boldsymbol{\varepsilon}^{el(Trial)}_{n+1}\right)$ the position of the stress state with respect to the evolved yield surfaces is checked. If plastic loading is the case, the trial state of stress calculated from the trial elastic strain is projected onto the current yield surface (see Fig.~\ref{fig:Returnmapping}). 
\begin{figure*}[htpb]
	  \begin{center}
	\includegraphics[scale=0.60]{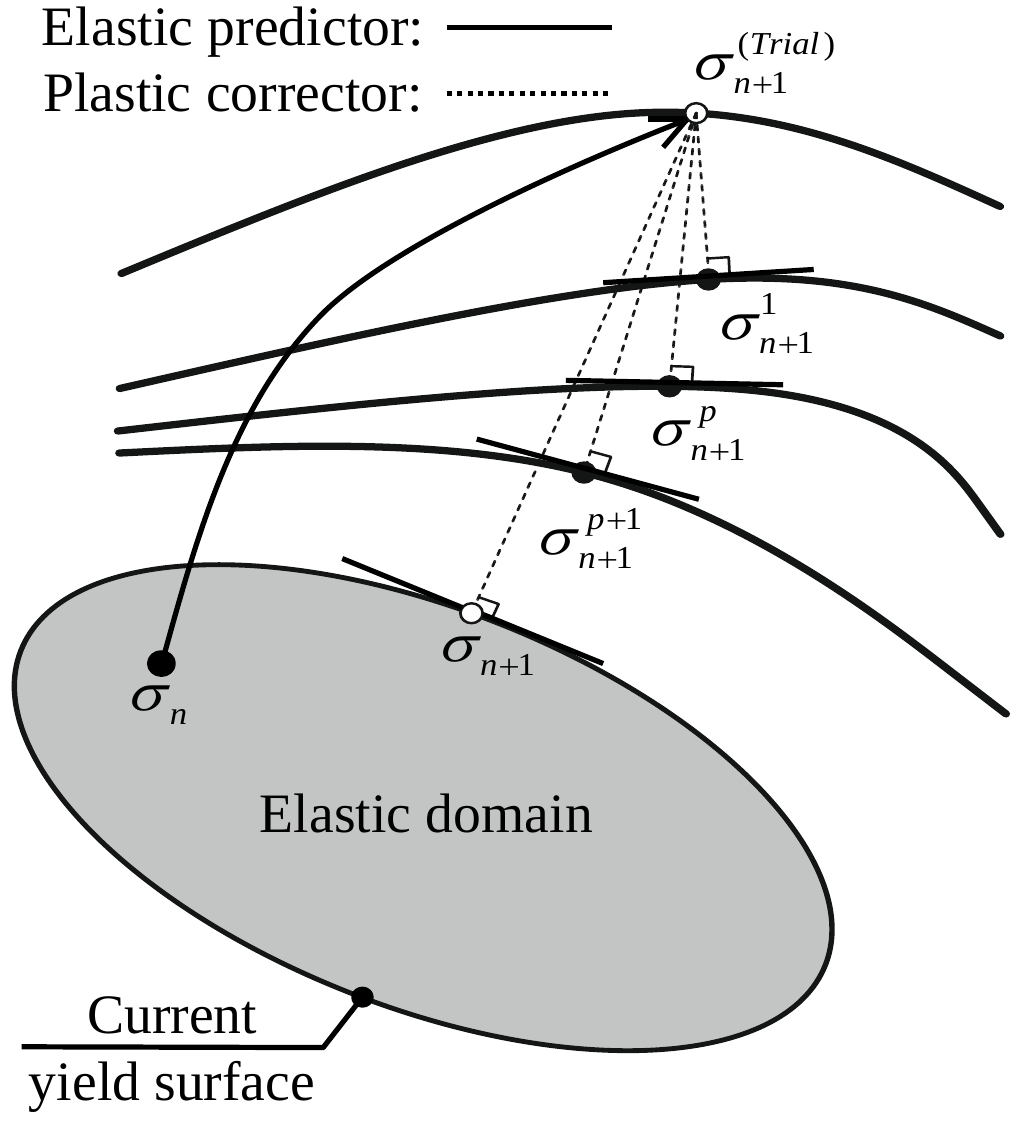}
	\caption{\label{fig:Returnmapping} Schematic representation of the concept of elastic predictor/plastic corrector in the context of the closest point projection iteration algorithm, based on Ref.\protect\cite{SimoHughes98}.} 
	  \end{center}
\end{figure*}

The plastic strain and respective internal variable(s) also need to be initialized as follows:
\begin{equation}
     \label{eq:Plasalpha}
     \boldsymbol{\varepsilon}^{pl(Trial)}_{n+1}=\boldsymbol{\varepsilon}^{pl(k=0)}_{n+1}=\boldsymbol{\varepsilon}^{pl}_{n}, \qquad \alpha_{l,n+1}^{(Trial)}=\alpha_{l,n+1}^{(k=0)}=\alpha_{l,n}~. 
\end{equation}
From a theoretical point of view, the trial state of elastic strain is achieved by suppressing the evolution of plastic flow within the time increment. As a consequence,~$\boldsymbol{\varepsilon}^{el(Trial)}_{n+1}$ and all corresponding state variables in the trial state read as~\cite{SimoHughes98}: 
\begin{align}
     \label{eq:TrialStat}
\boldsymbol{\varepsilon}^{el(Trial)}_{n+1}&=\left(\boldsymbol{\varepsilon}_{n+1}-\boldsymbol{\varepsilon}^{pl(Trial)}_{n+1}-\boldsymbol{\varepsilon}_{n+1}^{\omega}-\sum_{i=1}^{n}\boldsymbol{\varepsilon}_{i,n+1}^{ve(k)}-\sum_{j=1}^{m}\boldsymbol{\varepsilon}_{j,n+1}^{ms(k)}\right),     \\
\boldsymbol{\sigma}_{n+1}^{(Trial)}       &=\left({\bf C}_{0,n+1}:\boldsymbol{\varepsilon}^{el(Trial)}_{n+1}\right),     \\
q_{l,n+1}^{(Trial)}                       &=q_{l}\left(\alpha_{l,n+1}^{(Trial)},\omega_{n+1}\right)=q_{l}(\alpha_{l,n},\omega_{n+1}),     \\
f_{l,n+1}^{(Trial)}                       &=f_{l}\left(\boldsymbol{\sigma}_{n+1}^{(Trial)},q_{l,n+1}^{(Trial)},\omega_{n+1}\right),     
\end{align}
subsequently, based on the last expression for the yield functions if all~$f_{l,n+1}^{(Trial)}\leq0$ the deformation is purely elastic and consistency parameters of all yield mechanisms do not change ($\Delta{\gamma}^{l}=0$~for~$l=\mathrm{R, T, L}$), but if at least for one of them~$f_{l,n+1}^{(Trial)}>0$, the time step is plastic and the plastic strain increment along with changes of the strain hardening variable(s) and changes in the consistency parameters have to be evaluated. Therefore, by defining 
\begin{align}
     \label{eq:PlasStat}
\boldsymbol{\varepsilon}^{el(p)}_{n+1}&=\left(\boldsymbol{\varepsilon}_{n+1}-\boldsymbol{\varepsilon}^{pl(p)}_{n+1}-\boldsymbol{\varepsilon}_{n+1}^{\omega}-\sum_{i=1}^{n}\boldsymbol{\varepsilon}_{i,n+1}^{ve(k)}-\sum_{j=1}^{m}\boldsymbol{\varepsilon}_{j,n+1}^{ms(k)}\right),     \\
\boldsymbol{\sigma}_{n+1}^{(p)}       &=\left({\bf C}_{0,n+1}:\boldsymbol{\varepsilon}^{el(p)}_{n+1}\right),     \\
q_{l,n+1}^{(p)}                       &= q_{l}\left(\alpha_{l,n+1}^{(p)},\omega_{n+1}\right),     \\
f_{l,n+1}^{(p)}&=f_{l}\left(\boldsymbol{\sigma}_{n+1}^{(p)},q_{l,n+1}^{(p)},\omega_{n+1}\right),     
\end{align}
and by means of an iterative return-mapping algorithm (i.e., plastic corrector) incremental plastic deformations, hardening variable(s), and consistency parameter(s) associated with the active yield surface(s) are obtained. Note that superscripts~$p$ in Eq.~(\ref{eq:PlasStat}) refer to the~$p^{th}$ iteration of the return mapping algorithm. Consequently, iterative formulations concerning update of the state variables and consistency parameters are
\begin{align}
     \label{eq:AllPlasIncrem}
\boldsymbol{\varepsilon}^{pl(p+1)}_{n+1}&=\boldsymbol{\varepsilon}^{pl(p)}_{n+1}+\Delta{\boldsymbol{\varepsilon}^{pl(p)}_{n+1}},     \\
\alpha_{l,n+1}^{(p+1)}     &=\alpha_{l,n+1}^{(p)}+\Delta{\alpha_{l,n+1}^{(p)}},     \\
\Delta{\gamma}^{l(p+1)}_{n+1}&=\Delta{\gamma}^{l(p)}_{n+1}+\delta{\Delta{\gamma}^{l(p)}_{n+1}},~~~~~\mathrm{for ~all}~l\in\mathrm{\mathbb{S}}_{act}~.      
\end{align}
$\Delta{\gamma}^{l}_{n+1}$ denotes the non-rate form of the consistency parameter, whereas~$\delta\Delta{\gamma}^{l}_{n+1}$ signifies the corresponding increment. At the end of the return mapping iteration,~$(p+1)$ is set to~$(p)$ and the convergence of the iteration is checked. A plastic residuum vector, consisting of components related to the iterative evolution of the plastic strain tensor and of hardening variable(s):
\begin{equation}
     \label{eq:PlasResidu}
\boldsymbol{R}_{n+1}^{pl(p)}:=\left\{
     \begin{array}{cc}
-\boldsymbol{\varepsilon}^{pl(p)}_{n+1}+\boldsymbol{\varepsilon}^{pl}_{n} \\
-\boldsymbol{\alpha}_{n+1}^{(p)}+\boldsymbol{\alpha}_{n} \\
     \end{array} \right\}+\sum_{l~\in~\mathrm{\mathbb{S}}_{act}}^{r}\Delta{\gamma}^{l(p)}_{n+1}\left\{
          \begin{array}{cc}
\partial_{\boldsymbol{\sigma}}f_{l}(\boldsymbol{\sigma}^{(p)}_{n+1},\alpha_{l,n+1}^{(p)},\omega_{n+1})     \\ 
\partial_{\boldsymbol{q}}f_{l}(\boldsymbol{\sigma}^{(p)}_{n+1},\alpha_{l,n+1}^{(p)},\omega_{n+1})
          \end{array} \right\}
\end{equation}
is computed for the convergence check. Now the values of all active yield functions are recalculated using the updated stress tensor and hardening variable(s) via Eq.~(\ref{eq:YFun}). If 
\begin{equation}
     \label{eq:PlastCond}
f_{l,n+1}^{(p)}:=f_{l}(\boldsymbol{\sigma}^{(p)}_{n+1},\alpha_{l,n+1}^{(p)},\omega_{n+1})<\mathrm{TOL}_{1} ~\mathrm{for~all~}~l\in\mathrm{\mathbb{S}}_{act}~\mathrm{, and}~\left\|\boldsymbol{R}_{n+1}^{pl(p)}\right\|<\mathrm{TOL}_{2},   
\end{equation}
is not fully fulfilled, the procedure loops to the next iteration. Otherwise, the current iteration has converged and consequently, the definite increments of the plastic strain tensor, as well as the hardening variable(s) are obtained. Now all state variables relevant to the plastic part of the total strain are updated to the new values at the end of the time step, namely
\begin{equation}
     \label{eq:PlasUpdat}
\boldsymbol{\varepsilon}^{pl}_{n+1}= \boldsymbol{\varepsilon}^{pl}_{n}+\Delta{\boldsymbol{\varepsilon}^{pl}_{n+1}}, \qquad  \alpha_{l,n+1}                     = \alpha_{l,n}+\Delta{\alpha_{l,n+1}},~~~~~\mathrm{for ~all}~l\in\mathrm{\mathbb{S}}_{act}~. 
\end{equation}
 \item \textbf{Stage 3: Stress calculation} Now the driving stress in the current iteration of the general additive decomposition scheme can be calculated. Taking the strain components and the total strain tensor at the end of the increment, the elastic contribution is
\begin{equation}
     \label{eq:ElasticE}
     \boldsymbol{\varepsilon}^{el(k)}_{n+1}=
  \begin{dcases}
\boldsymbol{\varepsilon}_{n+1}-\boldsymbol{\varepsilon}^{pl(k)}_{n+1}-\boldsymbol{\varepsilon}_{n+1}^{\omega}-\sum_{i=1}^{n}\boldsymbol{\varepsilon}_{i,n+1}^{ve(k)}-\sum_{j=1}^{m}\boldsymbol{\varepsilon}_{j,n+1}^{ms(k)} & \mbox{, if}~   \boldsymbol{\varepsilon}^{pl(k)}_{n+1}\neq0,     \\
\boldsymbol{\varepsilon}_{n+1}-\boldsymbol{\varepsilon}_{n+1}^{\omega}-\sum_{i=1}^{n}\boldsymbol{\varepsilon}_{i,n+1}^{ve(k)}-\sum_{j=1}^{m}\boldsymbol{\varepsilon}_{j,n+1}^{ms(k)} & \mbox{, if}~   \boldsymbol{\varepsilon}^{pl(k)}_{n+1}=0.     \\
  \end{dcases}
\end{equation}
With the calculated elastic strain, the stress tensor at the~$\left(k\right)^{th}$ iteration of the stress update increment is
\begin{equation}
     \label{eq:Stressk}
\boldsymbol{\sigma}^{(k)}_{n+1}=\left({\bf C}_{0,n+1}:\boldsymbol{\varepsilon}^{el(k)}_{n+1}\right). 
\end{equation}

In a next step, the tangent operator for the whole serial model has to be given. For this reason, all individual Jacobians from the 3-4 deformation mechanisms are evaluated separately and assembled to a total algorithmic tangent operator \cite{HanHelnI2003,HanHelnII2003,Fortino09}. Since this step is crucial for the convergence of the entire implementation, a detailed explanation is given in the \ref{Sec:TangentOpt}. With the total tangent operator~$\boldsymbol{\mathrm{C}}^{T}_{n+1}$, the iterative change of the total stress is calculated by the expression
\begin{equation}     
     \label{eq:DeltaS}
\Delta\boldsymbol{\sigma}_{n+1}^{(k)}=-\boldsymbol{\mathrm{C}}^{T}_{n+1}:\left(\boldsymbol{\mathrm{R}}^{el}+\sum_{i=1}^{n}\boldsymbol{\mathrm{R}}_{i}^{ve}+\sum_{j=1}^{m}\boldsymbol{\mathrm{R}}_{j}^{ms}\right)_{n+1}^{(k)}.
\end{equation}
The residual vectors belonging to each deformation mechanism are defined as: 
\begin{align}
     \label{eq:ElasResVec}
\boldsymbol{\mathrm{R}}^{el(k)}&=     
  \begin{dcases}
{\bf C}_{0,n+1}^{-1}:\boldsymbol{\sigma}_{n+1}^{(k)}-\left(\boldsymbol{\varepsilon}_{n+1}-\boldsymbol{\varepsilon}^{pl(k)}_{n+1}-\boldsymbol{\varepsilon}_{n+1}^{\omega}-\sum_{i=1}^{n}\boldsymbol{\varepsilon}_{i,n+1}^{ve(k)}-\sum_{j=1}^{m}\boldsymbol{\varepsilon}_{j,n+1}^{ms(k)}\right) & \mbox{, if}~   \boldsymbol{\varepsilon}^{pl(k)}_{n+1}\neq0,     \\
{\bf C}_{0,n+1}^{-1}:\boldsymbol{\sigma}_{n+1}^{(k)}-\left(\boldsymbol{\varepsilon}_{n+1}-\boldsymbol{\varepsilon}_{n+1}^{\omega}-\sum_{i=1}^{n}\boldsymbol{\varepsilon}_{i,n+1}^{ve(k)}-\sum_{j=1}^{m}\boldsymbol{\varepsilon}_{j,n+1}^{ms(k)}\right) & \mbox{, if}~   \boldsymbol{\varepsilon}^{pl(k)}_{n+1}=0,      \\
  \end{dcases} \\
\label{eq:ViscResVec}
\boldsymbol{\mathrm{R}}_{i}^{ve(k)}&=\left\{\boldsymbol{\varepsilon}_{i,n}^{ve}~\mathrm{exp}\left(-\xi_{i}\right)+\mathrm{\mathbb{T}}_{n}^{ve}\left(\xi_{i}\right){\bf C}_{i,n}^{-1}:\boldsymbol{\sigma}_{n}+\mathrm{\mathbb{T}}_{n+1}^{ve}\left(\xi_{i}\right){\bf C}_{i,n+1}^{-1}:\boldsymbol{\sigma}_{n+1}^{(k)}-\boldsymbol{\varepsilon}_{i,n+1}^{ve(k)}\right\},   \\    
\label{eq:MechsResVec}
\boldsymbol{\mathrm{R}}_{j}^{ms(k)}&=\left\{\boldsymbol{\varepsilon}_{j,n}^{ms}~\mathrm{exp}\left(-\xi_{j}\right)+\mathrm{\mathbb{T}}_{n}^{ms}\left(\xi_{j}\right){\bf C}_{j,n}^{-1}:\boldsymbol{\sigma}_{n}+\mathrm{\mathbb{T}}_{n+1}^{ms}\left(\xi_{j}\right){\bf C}_{j,n+1}^{-1}:\boldsymbol{\sigma}_{n+1}^{(k)}-\boldsymbol{\varepsilon}_{j,n+1}^{ms(k)}\right\}.    
\end{align}
\item \textbf{Stage 4: Visco-elastic and mechano-sorptive deformation} Based on the pre-calculated incremental stress tensor Eq.~(\ref{eq:DeltaS}) and all decoupled residual vectors related to all individual deformation modes Eqs.~(\ref{eq:ElasResVec}-\ref{eq:MechsResVec}), the change of visco-elastic strains in conjunction with the variation of mechano-sorptive deformations for the current iteration of the stress update algorithm are determined. Accordingly, the visco-elastic and mechano-sorptive creep strain increments~$\Delta\boldsymbol{\varepsilon}_{i,n+1}^{ve(k)}$ and~$\Delta\boldsymbol{\varepsilon}_{j,n+1}^{ms(k)}$ are evaluated via the following relationships
\begin{align}
     \label{eq:DeltaVeMs}
     \nonumber
\Delta\boldsymbol{\varepsilon}_{i,n+1}^{ve(k)}&=\boldsymbol{\mathrm{R}}_{i}^{ve(k)}~+\mathrm{\mathbb{T}}_{n+1}^{ve}\left(\xi_{i}\right){\bf C}_{i,n+1}^{-1}:\Delta\boldsymbol{\sigma}_{n+1}^{(k)}~~~~~~~~\mbox{, for}~i=\left(1,...,n\right),     \\
\Delta\boldsymbol{\varepsilon}_{j,n+1}^{ms(k)}&=\boldsymbol{\mathrm{R}}_{j}^{ms(k)}+\mathrm{\mathbb{T}}_{n+1}^{ms}\left(\xi_{j}\right){\bf C}_{j,n+1}^{-1}:\Delta\boldsymbol{\sigma}_{n+1}^{(k)}~~~~~~~~\mbox{, for}~j=\left(1,...,m\right),          
\end{align}
that are derived based on the definition of respective algorithmic operators (see \ref{Sec:TangentOpt}). Consecutively, all state variables for all element-wise visco-elastic and mechano-sorptive creep deformations are updated at the end of the ongoing iteration using the expression
\begin{equation}
     \label{eq:VeMsUpdate}
\boldsymbol{\varepsilon}_{i,n+1}^{ve(k+1)}=\boldsymbol{\varepsilon}_{i,n+1}^{ve(k)}+\Delta\boldsymbol{\varepsilon}_{i,n+1}^{ve(k)}, \qquad \boldsymbol{\varepsilon}_{j,n+1}^{ms(k+1)}=\boldsymbol{\varepsilon}_{j,n+1}^{ms(k)}+\Delta\boldsymbol{\varepsilon}_{j,n+1}^{ms(k)}.   
\end{equation}
The iterative procedure is completed by comparing a generalized residual vector~${\bf{R}}_{n+1}^{(k+1)}$
\begin{equation}
     \label{eq:GenResidu}
{\bf{R}}_{n+1}^{(k+1)}=\left\{\boldsymbol{\mathrm{R}}_{n+1}^{el(k+1)}~\boldsymbol{\mathrm{R}}_{i,n+1}^{ve(k+1)}\cdots\boldsymbol{\mathrm{R}}_{j,n+1}^{ms(k+1)}\cdots\right\}^{T},     
\end{equation}
composed of all recomputed deformation-based residual vectors Eqs.~(\ref{eq:ElasResVec}-\ref{eq:MechsResVec}), or more precisely its norm up to a tolerance level. If~$\left\|{\bf{R}}_{n+1}^{(k+1)}\right\|\leq\mbox{TOL}_{3}$ the iterative scheme is converged and the solution obtained in the~$(k+1)^{th}$ iteration is regarded as the final response of the current increment. Otherwise, if~$\left\|{\bf{R}}_{n+1}^{(k+1)}\right\|>\mbox{TOL}_{3}$,~$(k+1)$ is changed to~$(k)$ and stages 2-4 are repeated. After convergence, all values of the state variables related to the visco-elastic and mechano-sorptive strains, as well as the elastic deformation Eq.~(\ref{eq:ElasticE}) are substituted by the updated ones:
\begin{equation}
     \label{eq:VeMsFinUp}
\boldsymbol{\varepsilon}_{i,n+1}^{ve}=\boldsymbol{\varepsilon}_{i,n}^{ve}+\Delta\boldsymbol{\varepsilon}_{i,n+1}^{ve},  \qquad \boldsymbol{\varepsilon}_{j,n+1}^{ms}=\boldsymbol{\varepsilon}_{j,n}^{ms}+\Delta\boldsymbol{\varepsilon}_{j,n+1}^{ms}.   
\end{equation}
These updated state values will be considered as initial values for the next time step.
\end{enumerate}
\nomenclature{$\boldsymbol{\varepsilon}^{el(Trial)/pl(Trial)}_{n+1}$}{Trial elastic/Trial plastic strain tensor (at~$t_{n+1}$)}
\nomenclature{$\alpha_{l,n+1}^{(Trial)}$}{Trial internal hardening variable (at~$t_{n+1}$),~$l=$R, T, L}
\nomenclature{$\boldsymbol{\varepsilon}_{i/j,n+1}^{ve/ms(k)}$}{$i/j^{th}$ visco-elastic/mechano-sorptive strain tensor after $\left(k\right)^{th}$ iteration of the strain decomposition algorithm (at~$t_{n+1}$)}
\nomenclature{$\boldsymbol{\sigma}_{n+1}^{(Trial)}$}{Trial total stress tensor (at~$t_{n+1}$)}
\nomenclature{$f_{l,n+1}^{(Trial)}/q_{l,n+1}^{(Trial)}$}{Trial yield function/Trial hardening function value (at~$t_{n+1}$),~$l=$R, T, L}
\nomenclature{$\Delta{\gamma}^{l(p)}_{n+1}/\delta\Delta{\gamma}^{l(p)}_{n+1}$}{Non-rate form/Change of the consistency parameter after~$\left(p\right)^{th}$ iteration of return-mapping,~$l=$R, T, L}
\nomenclature{$\boldsymbol{\varepsilon}^{el(p)/pl(p)}_{n+1}$}{Elastic/Plastic strain tensor after~$\left(p\right)^{th}$ iteration of return-mapping (at~$t_{n+1}$)}
\nomenclature{$\boldsymbol{\sigma}_{n+1}^{(p)}$}{Total stress tensor after~$\left(p\right)^{th}$ iteration of return-mapping (at~$t_{n+1}$)}
\nomenclature{$\alpha_{l,n+1}^{(p)}$}{Internal hardening variable after~$\left(p\right)^{th}$ iteration of return mapping (at~$t_{n+1}$),~$l=$R, T, L}
\nomenclature{$f_{l,n+1}^{(p)}/q_{l,n+1}^{(p)}$}{Yield function/Hardening function value after~$\left(p\right)^{th}$ iteration (at~$t_{n+1}$),~$l=$R, T, L}
\nomenclature{$\Delta{\boldsymbol{\varepsilon}^{pl(p)}_{n+1}}/\Delta{\alpha_{l,n+1}^{(p)}}$}{Change of the plastic strain tensor/Change of the internal hardening variable after~$\left(p\right)^{th}$ iteration of return-mapping}
\nomenclature{$\boldsymbol{R}_{n+1}^{pl(p)}$}{Plastic residual vector after~$\left(p\right)^{th}$ iteration of return-mapping (at~$t_{n+1}$)}
\nomenclature{$\partial_{\boldsymbol{q}}f_{l}$}{Hardening strain flow direction}
\nomenclature{$\Delta{\boldsymbol{\varepsilon}^{pl}_{n+1}}$}{Increment of the plastic strain tensor from~$t_{n}$ to~$t_{n+1}$}
\nomenclature{$\Delta{\alpha_{l,n+1}}$}{Increment of the internal hardening variable from~$t_{n}$ to~$t_{n+1}$,~$l=$R, T, L}
\nomenclature{$\Delta\boldsymbol{\sigma}_{n+1}^{(k)}$}{Change of the total stress tensor after~$\left(k\right)^{th}$ iteration}
\nomenclature{$\boldsymbol{\mathrm{R}}^{el}, \boldsymbol{\mathrm{R}}_{i/j}^{ve/ms}$}{Elastic,~$i/j^{th}$ visco-elastic/mechano-sorptive residual vector}
\nomenclature{$\Delta\boldsymbol{\varepsilon}_{i/j,n+1}^{ve/ms(k)}$}{Change of the~$i/j^{th}$ visco-elastic/mechano-sorptive strain tensor after~$k^{th}$ iteration}
\nomenclature{${\bf{R}}_{n+1}^{(k+1)}$}{Generalized residual vector after~$\left(k+1\right)^{th}$ iteration of the strain decomposition algorithm}
\nomenclature{$\Delta\boldsymbol{\varepsilon}_{i/j,n+1}^{ve/ms}$}{Increment of~$i/j^{th}$ visco-elastic/mechano-sorptive strain tensor from~$t_{n}$ to~$t_{n+1}$}
\section{Verification of the material model}\label{Sec:Verif}
\begin{figure*}[htpb]
     \begin{center}
	\includegraphics[scale=0.65]{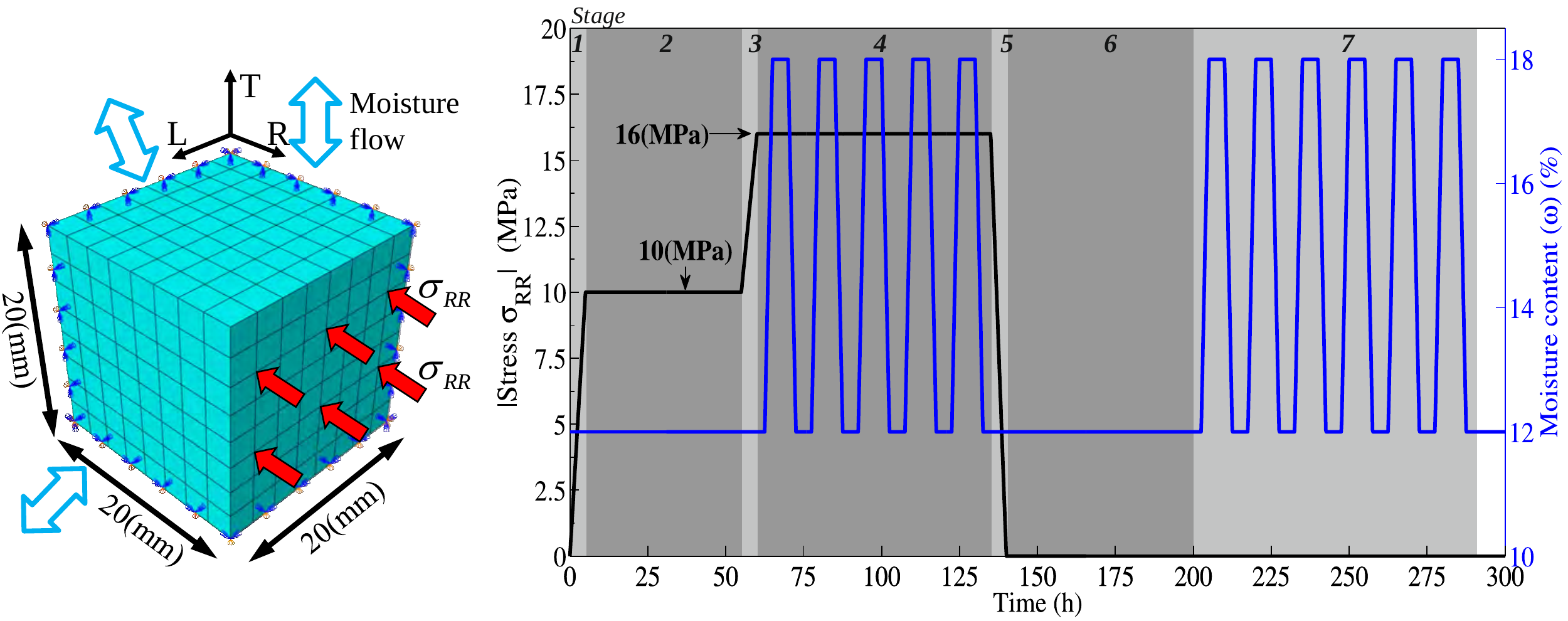}
    \caption{Left: Geometry and finite element model of the wood cubic specimen. Right: Amplitudes of exerted radial compressive stress and moisture distribution on the wood block sample. (Example 1)}
   \label{fig:MeshAmpl_6}
   \end{center}
\end{figure*}

A set of three numerical examples is calculated that verifies the capability and efficiency of the proposed 3D constitutive model for the prediction of realistic behavior for short-term and long-term responses under combined moisture and mechanical loading. The examples are selected in such a way that different deformation components act in an isolated and combined way under uni- and multi-axial loading, as well as for restrained swelling. All examples use the material properties given in the \ref{Sec:Tables} for European beech (\emph{Fagus sylvatica} L.). Furthermore, this section is complemented with an experimental verification.
\paragraph{Example 1 - Uni-axial loading:~} In a first simple example, a cubic sample with edge length 40mm is studied. The model uses three confining symmetry planes, reducing the edge length to 20mm (see Fig.~\ref{fig:MeshAmpl_6} (left)). Hence it is allowed to swell or shrink freely during moisture variations and hygro-expansions do not lead to residual stresses. Since the dimensions are quite small compared to the ones of a stem, the curvature of growth rings is ignored by assigning an orthotropic material behavior to the system that is defined in a Cartesian coordinate system with axes aligned along the cube edges. The geometry is discretized by 512 quadratic brick elements (C3D20) and loaded by a uniform compression in radial direction. Simultaneously a homogeneous moisture distribution can be applied (see Fig.~\ref{fig:MeshAmpl_6} (right)). The pressure and moisture content~$\omega$ are chosen in a way that all potential components of the total strain are addressed and can be distinguished clearly during the 7 stages:
\begin{itemize}
	\item \textbf{Stage 1} 0-5h: The sample is at standard climate, i.e., 65\% RH resulting in~$\omega_{1}$=12\%. The pressure is ramped up to 10MPa during 5 hours,
	\item \textbf{Stage 2} 5-55h: for the next 50 hours all conditions are kept constant,
	\item \textbf{Stage 3} 55-60h: within the next 5 hours, pressure is increased from 10 to 16MPa, while~$\omega_{1}=12\%=$const.,
	\item \textbf{Stage 4} 60-135h: at constant load,~$\omega$ is going through 5 cycles of wetting and drying, each lasting 15 hours. For 2.5h~$\omega_{1}=12\%$, then ramped up to~$\omega_{2}=18\%$ within 2.5h, held constant for 5 hours, ramped down from~$\omega_{2}$ to~$\omega_{1}$ for 2.5h, held constant for 2.5h, a.s.o..
	\item \textbf{Stage 5} 135-140h: At~$\omega_{1}$ the load is completely removed during 5h and will remain this way for the rest of the simulation.
	\item \textbf{Stage 6} 140-200h: For the next 60 hours all conditions are kept constant.
	\item \textbf{Stage 7} 200-290h: Finally six more moisture cycles like in Stage 4, but without mechanical loading are imposed.
\end{itemize}

\begin{figure*}[htpb]
     \begin{center}
            \includegraphics[width=1.00\textwidth]{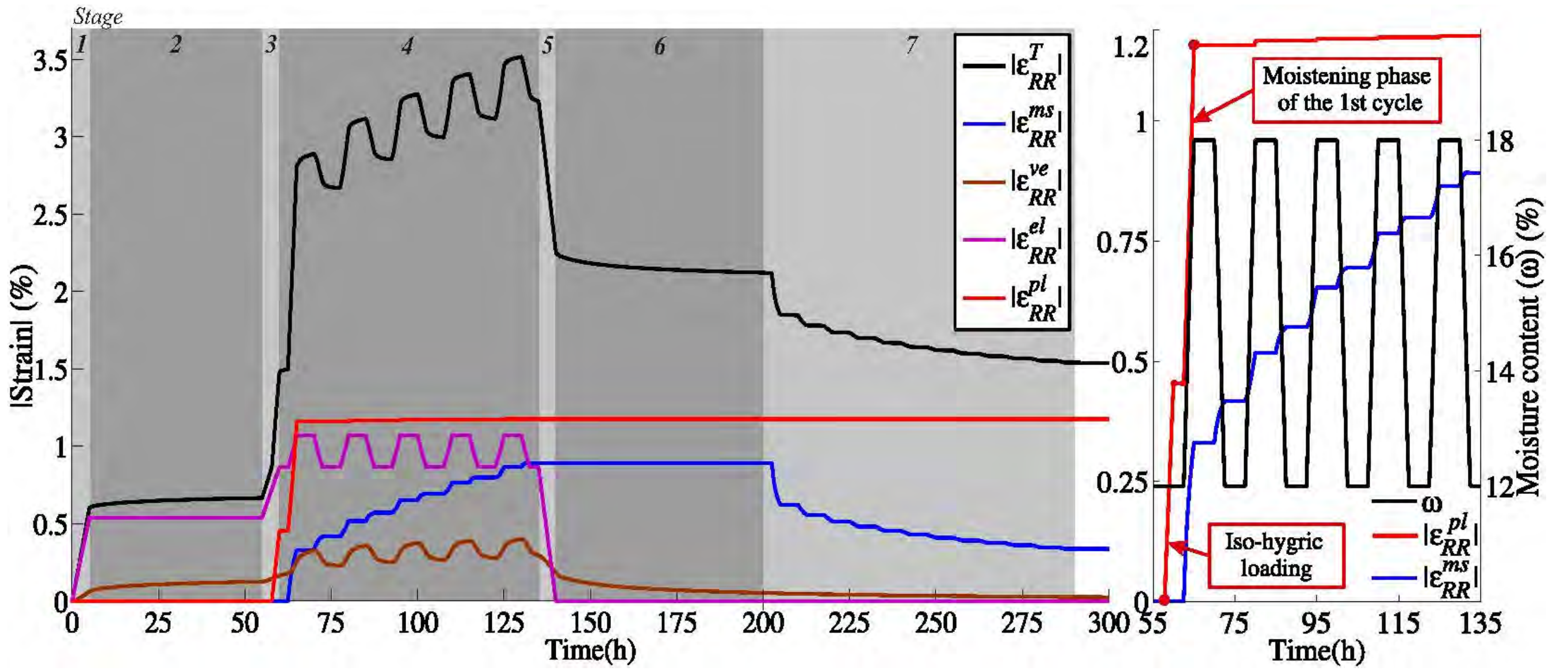}      
   \caption{Left: Time evolution of all constituents of the total strain along radial direction. Right: Development of radial irrecoverable plastic strain and mechano-sorptive creep. (Example 1)}
   \label{fig:TotStr_R_7}
   \end{center}
\end{figure*}

The resulting strain along the radial direction for the loading in the 7 stages is illustrated in Fig.~\ref{fig:TotStr_R_7}. In order to observe the development of all partial strains in an easy to interpret way, the hygro-expansion strain is subtracted from the total strain and the remaining value, called \emph{swelling/shrinkage-modified} strain, is shown in Fig.~\ref{fig:TotStr_R_7} (left). For the different stages we observe:
\begin{itemize}
	\item \textbf{Stage 1}     0-5h: Linear elastic behavior dominates.
	\item \textbf{Stage 2}    5-55h: Increase in visco-elastic creep is noticed.
	\item \textbf{Stage 3}   55-60h: Plastic deformation starts as the compressive strength of the material along radial direction (i.e., -13.4MPa) is reached. The radial yield criterion is activated, material starts to flow, and the irrecoverable plastic deformation evolves as the material simultaneously hardens until the yield surface reaches 16MPa.
	\item \textbf{Stage 4}  60-135h: During the first moisture increase from~$\omega_{1}$=12(\%) to~$\omega_{2}$=18(\%), the material strength values degrade and consequently further plastic deformation, even under fixed external loading, evolves (see Fig.~\ref{fig:TotStr_R_7} (right)). The repetition of moistening in the following cycles just leads to a comparably small increase in the irreversible part of the total strain. The increase of strain during the moisture cycles is therefore dominated by mechano-sorptive creep. Note that the general response of the material model representing the simultaneous interaction of fixed level of loading and varying moisture content shows a good agreement with experimental observations of mechano-sorptive creep reported by Refs.~\cite{Toratti92,Houska95,ToratSven02}. It can be noticed that both moistening and de-moistening leads to the increase of mechano-sorptive deformation and generally, the material behavior demonstrates an ascending trend which implies that mechano-sorptive creep is accumulated over time.
	\item \textbf{Stage 5}  135-140h: During unloading, the instantaneous elastic response is immediately compensated.
	\item \textbf{Stage 6}  140-200h: Visco-elastic creep is partly recovered.
	\item \textbf{Stage 7}  200-290h: Partial recovery of mechano-sorptive creep.
\end{itemize}
It can be concluded that retrieval of the visco-elastic deformation requires adequate time, while the recovery of the mechano-sorption is achieved by moisture variation. As can be observed in Fig.~\ref{fig:TotStr_R_7} (left) even though the external pressure is removed, a remarkable amount of strain stays in the material. A part of this remaining deformation consists of time-dependent responses and another part is due to the occurrence of irreversible plastic deformation. 

\begin{figure*}[htpb]
	  \begin{center}
	\includegraphics[scale=0.500]{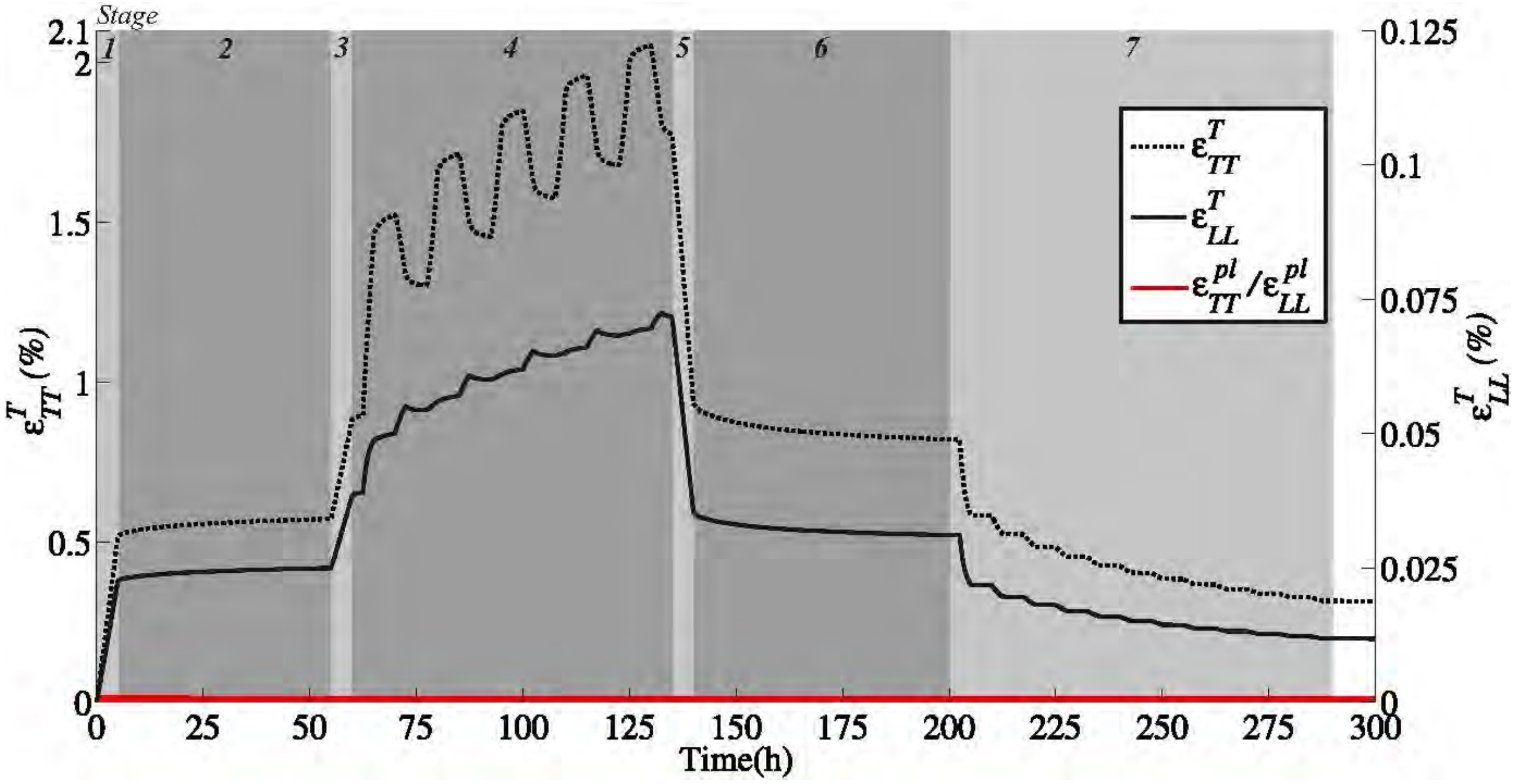}
	\caption{Evolution of swelling/shrinkage-modified strains along T- and L-direction. (Example 1)}
	\label{fig:Str_TL_10}
	  \end{center}
\end{figure*}

In addition, mechanical responses of the wood block along two other directions, i.e., tangential and longitudinal are also given in Fig.~\ref{fig:Str_TL_10}. Although no mechanical loading is applied along these two directions, significant lateral strains are generated. Contrary to the radial direction and as a consequence of the implemented multi-surface plasticity, where all failure mechanisms in each anatomic direction are independent, no plastic deformations can occur in T- and L-direction. Therefore all deformations are fully recoverable after time and moisture variations under zero external loading. The offset between the two directions can be explained by the fact that beech is much stiffer parallel than perpendicular to grain. Additionally, since the longitudinal entry of the mechano-sorptive compliance tensor was calculated differently than the other components (see Section~\ref{SSSec:Mechano}), the general mechano-sorptive response of the material model parallel to growth direction differs qualitatively from the corresponding behavior perpendicular to grain.

\begin{figure*}[htpb]
     \begin{center}
     \includegraphics[width=0.8\textwidth]{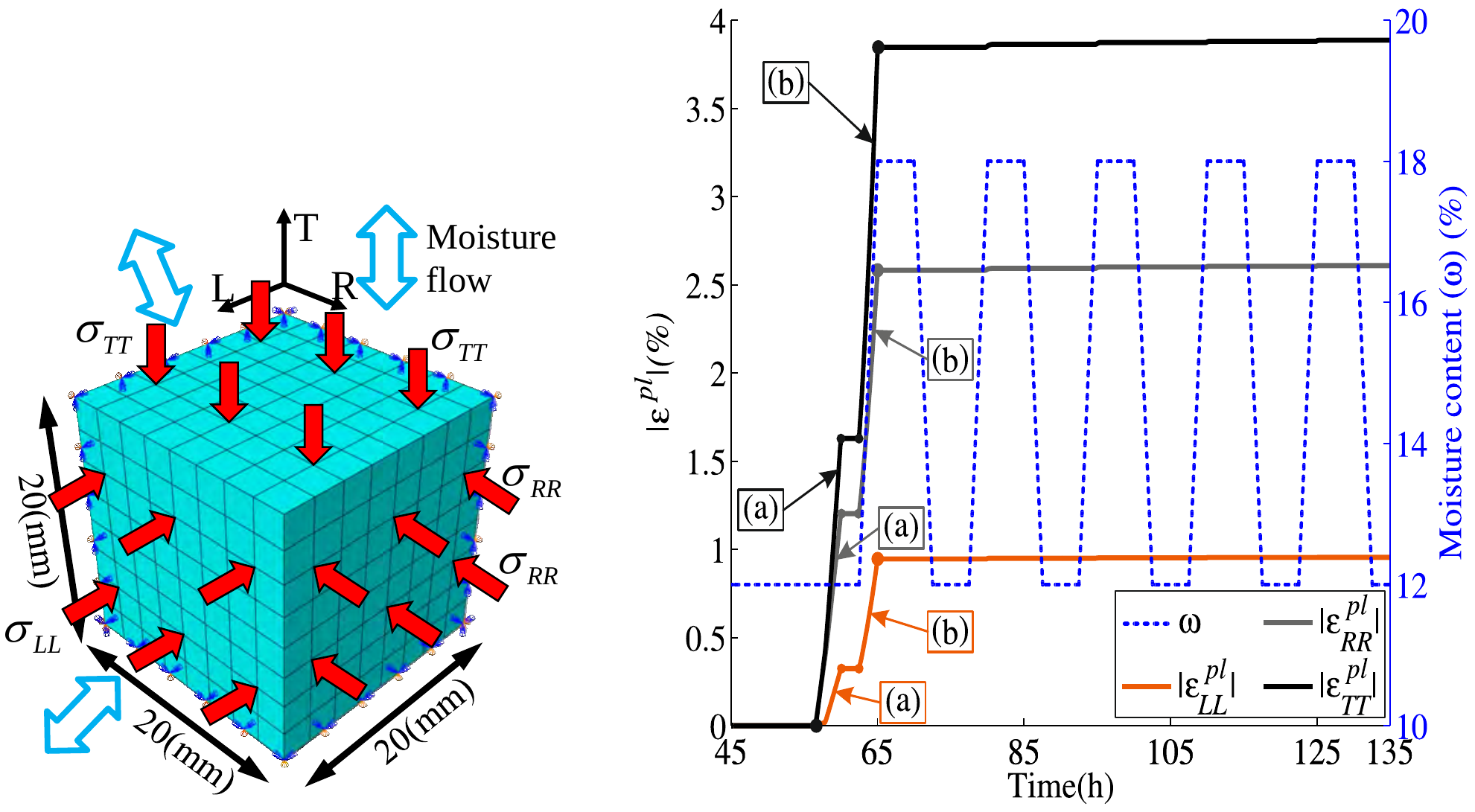}             
   \caption{Left: Illustration of the tri-axial state of stress. Right: Time evolution of irreversible plastic strains along all three anatomical directions. (a) and (b) denote iso-hygric loading and the moistening phase of the $1^{st}$ moisture cycle, respectively. (Example 2)}
   \label{fig:PlastStr_RTL_12}
   \end{center}
\end{figure*}

\paragraph{Remarks on convergence}
The classical \emph{Newton-Raphson} scheme is employed to solve non-linear problems within the framework of FE environment at which the material model has been implemented. Therefore, regarding the performance and the rate of convergence of the proposed computational algorithm, application of consistent tangent operator is of great significance. Convergence is evaluated in terms of residual control parameter ($R^{\alpha}$), which is calculated as the fraction of the largest residual in the equilibrium equation for the field displacement to the mean value of the conjugate force flux. Here, the convergence criterion is assumed as~$5\times10^{-3}$ which would appear rather stringent for engineering applications, but to achieve precise solutions to non-linear problems such a small tolerance is inevitable.  
Table~\ref{table:ConverNorm} in the \ref{Sec:Tables} shows the computational effort for some typical increments during moistening phase of the first moisture cycle for~$t=$62.5 to 65h. As it can be observed clearly from the data given in Table~\ref{table:ConverNorm}, the quadratic convergence rate of the implemented material model is obvious. At most three iterations are sufficient to meet the convergence tolerance in spite of increase in the moisture content resulting in the variation of model operator in every increment and appearance of further non-linear behavior. Although it was supposed that such a strict convergence tolerance imposes a very difficult condition to fulfill, via this numerical example it is illustrated that the convergence is satisfied after few number of iterations. Due to application of consistent Jacobian, the general robustness of the computational model is preserved.
\nomenclature{$R^{\alpha}$}{Residual control parameter}
\paragraph{Example 2 - Multi-axial loading:~} To evaluate the model behavior under arbitrary combinations of load with multiple yield surfaces activated, a tri-axial state of stress is imposed on the same specimen, but now with load in all anatomical directions (see Fig.~\ref{fig:PlastStr_RTL_12} (left)). Boundary conditions, as well as loading history are identical to the first example (see Fig.~\ref{fig:MeshAmpl_6}) only that the load in tangential direction is scaled by a factor of 0.4 and the one in longitudinal by a factor of 4.

\begin{figure*}[htpb]
     \begin{center}
     \includegraphics[width=1\textwidth]{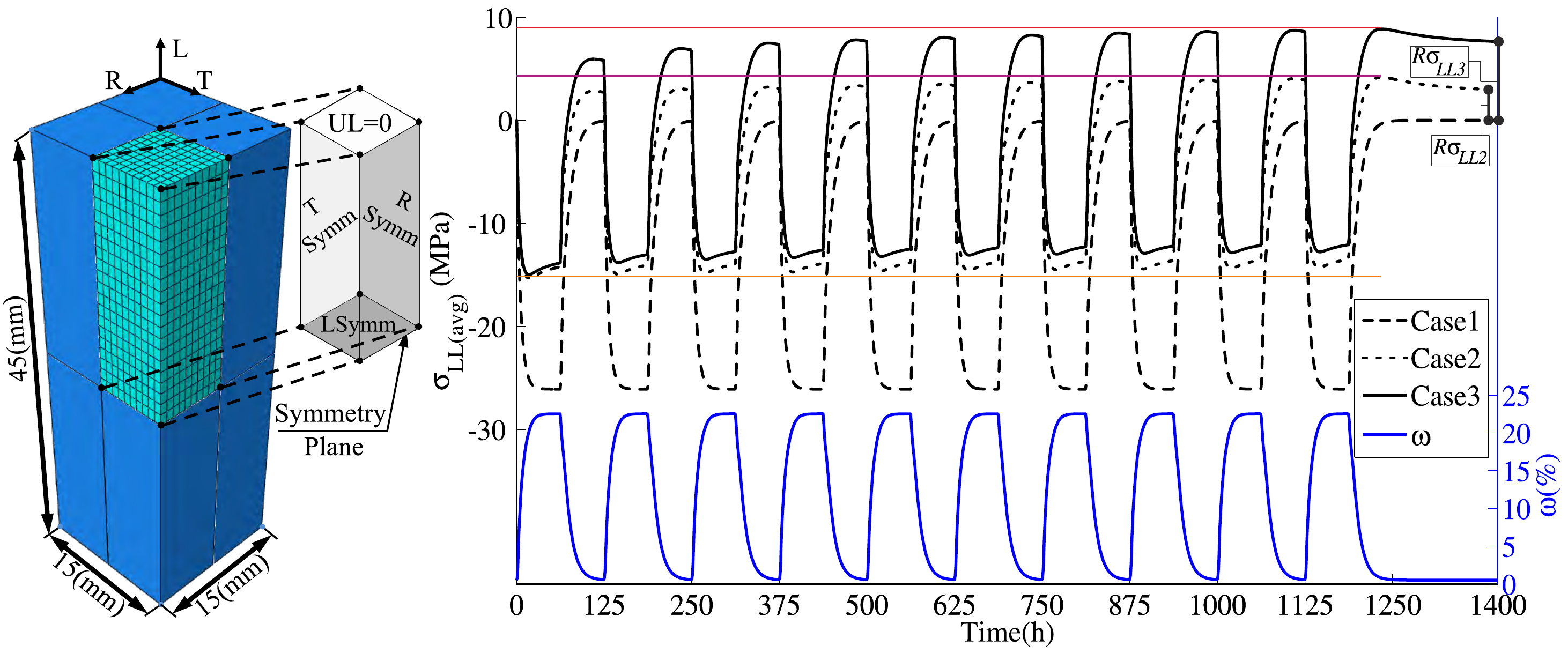}             
     \caption{Left: Finite element discretization of the rectangular prism specimen. Right: Time evolution of swelling pressure.~$R\sigma_{LL2}$ and~$R\sigma_{LL3}$ refer to the residual longitudinal stresses based on cases 2 and 3, respectively after $t=$1400h. (Example 3)}
     \label{fig:PrismMshStrss_10}
     \end{center}
\end{figure*}

For the second simulation the same interpretations as for the first example apply, concerning the evolution of all instant and time-dependent mechanical behavior. However, in contrast to example 1, all three yield mechanisms will be activated and consequently, irrecoverable plastic deformations evolve along all three anatomic directions simultaneously. The resulting plastic deformations are illustrated in Fig.~\ref{fig:PlastStr_RTL_12} (right). As visible, the tangential direction deforms more plastically, since it is the least stiff orientation. Furthermore, it should be noted that due to the application of a consistent tangent operator, the quadratic rate of convergence is preserved, even when multiple failure mechanisms are active.
\paragraph{Example 3 - Restrained swelling:~} As a last example, we simulate restrained swelling, as this is an experimentally studied loading case of practical relevance. We observe the evolution of the swelling pressure, as the rectangular sample of size 45mm$\times$15mm$\times$15mm (L$\times$R$\times$T) is exposed to a cycling moisture change (see Fig.~\ref{fig:PrismMshStrss_10}). This transient numerical simulation aims to assess the practical behavior of wood components in terms of moisture distribution, all possibly emerging deformation mechanisms, moisture-dependent plastification in particular, and associated stress fields, generated during moistening processes. Hence moisture transport is calculated based on the procedure outlined in Section~\ref{SSec:Moisture}. Like in the previous examples, we make use of the symmetric nature of the problem, and additionally to the three symmetry planes, we impose a non-moving boundary in L-direction at the top surface. The moisture content cycles from oven-dry condition ($\omega_{1}=0.5\%$) in 62.5h to a state with 95\% RH, in~$\omega_{2}=22.5\%$ close to the fiber saturation, only that moisture diffuses into the system from the exposed surfaces. Then, to observe the permanent deformations, the moisture content is again returning back to~$\omega_{1}=0.5\%$ for the next 62.5(h). To identify the long-term effects of history- and moisture-dependent deformation modes on the resulting strain and stress fields, the moistening/de-moistening cycle is repeated ten times. In addition, to provide more time for the recovery of the time-dependent deformation mechanisms, the analysis is run for further 150 hours.

In order to investigate the role of instantaneous and long-term responses on the mechanical behavior of wood specimen, this simulation is carried out for three different cases: \textit{(Case 1:)} Purely elastic material model with instantaneous responses. \textit{(Case 2:)} Material behavior with a rheological model consisting of history-dependent modes, i.e., visco-elasticity and mechano-sorption with no plasticity, \textit{(Case 3:)} Like case 2, but including plasticity. Fig.~\ref{fig:PrismMshStrss_10} (right) shows the resulting time evolution of swelling pressures as the mean value of the longitudinal stress in the symmetry plane (see Fig.~\ref{fig:PrismMshStrss_10} (left)).

As expected, increasing moisture content and swelling constraints result in compressive stresses. Their maximum is significantly higher for the case 1 than for the other two cases, convincingly demonstrating that a realistic stress analysis for wood should not be of purely elastic nature. On the basis of material models with time- and moisture-dependent behavior as well as plastification, i.e., cases 2,3 and in accordance with (Eqs.~(\ref{eq:ElasticE}) and~(\ref{eq:Stressk})), more terms are subtracted from the total strain value and subsequently, elastic strain and equivalently total stress are lower. In case 3, due to the occurrence of plastic deformation, this reduction in the elastic strain is even more pronounced. Here, even significant tensile stresses can be built up due to demoistening to overcome compressive time-dependent or plastic deformations. For case 3 these are the highest, and it is interesting to observe that due to repetition of moisture cycles, opposite to the fully recoverable time-dependent mechanisms, for the full model, these stresses even increase - a long-term effect that is not considered in any calculation for construction elements. By considering all three numerical examples, it can be observed that the developed material model can be applied to any arbitrary combination of mechanical loading, inhomogeneous moisture distribution and boundary condition. In what follows the experimental verification of the presented constitutive model is presented.
\paragraph{Experimental validation of an uncross-wise laminated sample:~}
To investigate the practical behavior of adhesively bonded wood elements under varying relative humidity, a three-layered uncross-wise laminated European beech specimen was exposed to moisture variations. Initially conditioned at 95\% RH layers were glued together using PRF (phenol resorcinol formaldehyde) adhesive. Subsequently, the composite sample was placed inside a climate box with varying RH comprising of three consecutive steps: 1) de-moistening from 95\% RH to 2\% RH; 2) re-moistening to 95\% RH; 3) another de-moistening to 2\% RH. At the end of each de-moistening cycle, the moisture-induced warping and sample dimensions were measured by means of a dial gauge. 

\begin{figure*}[htpb]
	  \begin{center}
	\includegraphics[scale=0.60]{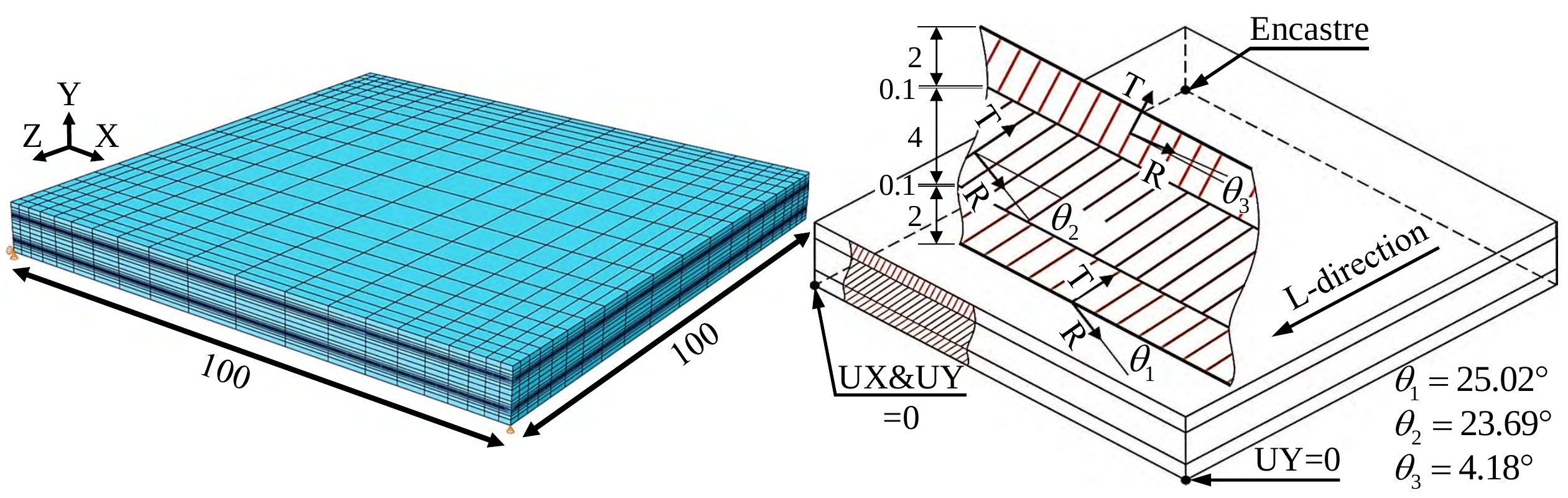}
	\caption{Left: Finite element model of the uncross-wise laminated sample. Right: Applied boundary conditions and Cartesian material coordinates. Indices refer to the layers. All dimensions are in mm.}
	\label{fig:PRFGeom_12}
	  \end{center}
\end{figure*}

The FEM model consists of five layers: three lamellae for wood substrates and two adhesive layers with a thickness of 0.1 mm (see Fig.~\ref{fig:PRFGeom_12}).

The variation of the RH inside the climate box is converted to the equivalent moisture content by means of the sorption isotherm curves for both European beech and PRF adhesive \cite{OliverSorption13}. These values are applied as time-dependent surface conditions to all external faces of the FE model during the moisture analysis. Using the described wood rheological model and a purely linear elastic isotropic behavior for the adhesive layers, the moisture-induced stress and the deformation fields are computed subsequently. Note that the adhesive properties are functions of moisture content as well with all parameters being summarized in Table~\ref{table:AdhPRF} in the \ref{Sec:Tables}. 

\begin{figure*}[htpb]
	  \begin{center}
	\includegraphics[scale=0.70]{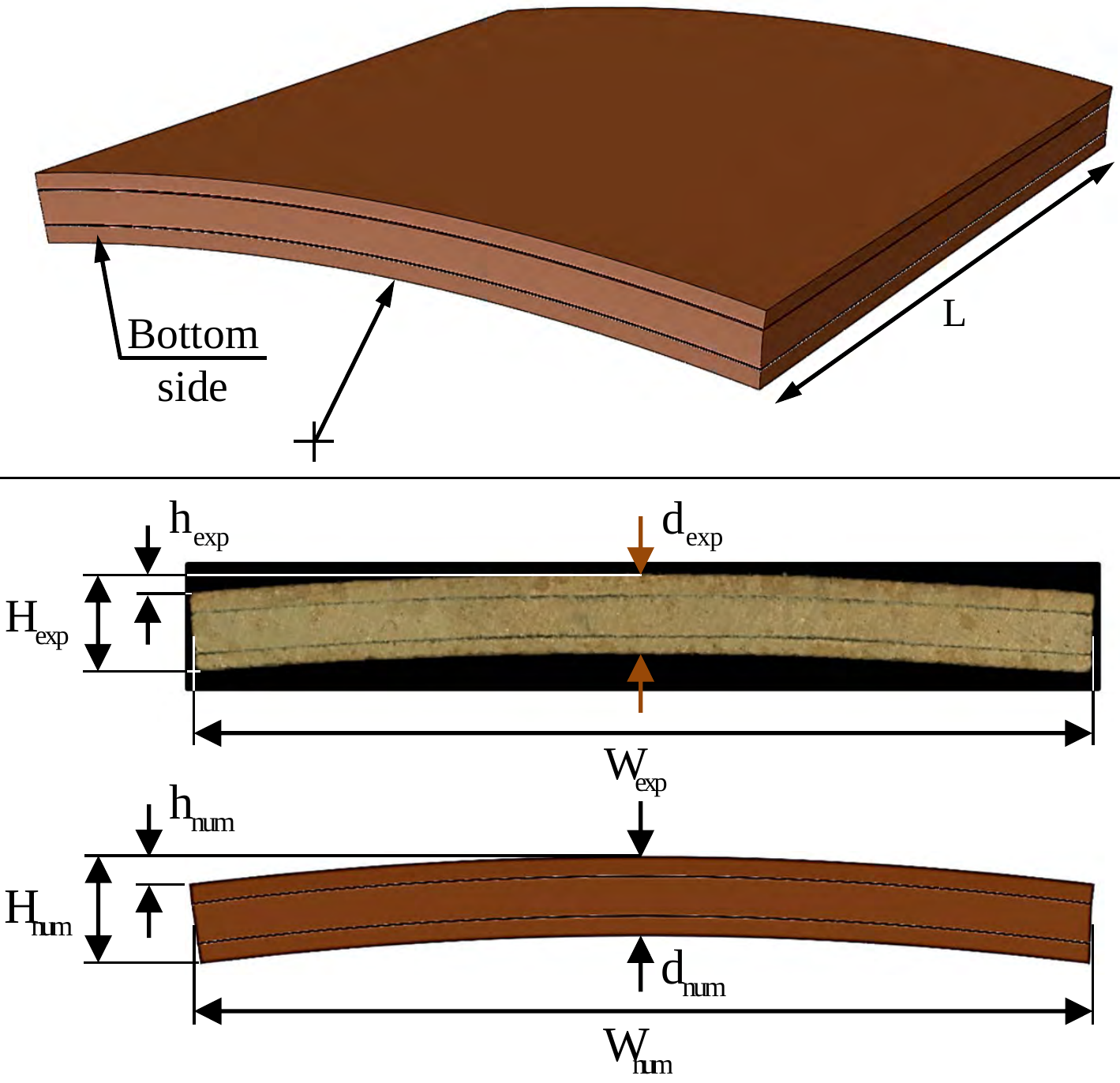}
	\caption{Deformed shape at 2\% RH. (Top) 3D view and (bottom) side view of experiment and simulation, naming the measurement points for quantitative comparisons in the Table~\ref{table:PRFExpNum}.}
	\label{fig:PRFDeform_13}
	  \end{center}
\end{figure*}

The comparison of simulation results with experimental observations gives excellent agreement, implying that the material model performs well for practical applications (see Fig.~\ref{fig:PRFDeform_13} and Table~\ref{table:PRFExpNum}). 

The quantitative comparison of moisture related deformations between experiment and simulation (Table~\ref{table:PRFExpNum}) gives excellent agreement considering that the utilized material properties are general values and do not consider the large variations common to wood. By having a more detailed look at the Table~\ref{table:PRFExpNum} it can be realized that the dimension ''H'', called here as cup deformation, has identically increased in experiment+simulation for the second de-moistening compared to the first one. This augmentation is due to time- and history-dependent deformation mechanisms. Note that even though in this example the difference in ''H'' is only 0.1 mm, it increases with the number of cycles and can for different configurations, like cross-lamination or hybrid components add to reaching critical values.
\section{Application example: Hybrid wood elements}\label{Sec:Example}
Combinations of different materials and wood species are a promising approach to overcome design limitations that result in extreme cross sections of structural elements made of softwood or in expensive solutions for load transfer from other elements perpendicular to grain, just to give two examples. The combination of beech and spruce has recently attracted significant attention due to a general technical approval for such a hybrid glulam element. It is interesting to note that neither plastic behavior, nor long-term effects did play a role in the approval procedure. To demonstrate the capability of the introduced approach, a hybrid glulam beam made of European beech and Norway spruce is considered. Both were adhesively joined at initial moisture content of 10\%. The multi-species beam is subjected to changing environmental conditions varying between 50\% RH and 90\% RH. Although European beech and Norway spruce show different moisture sorption isotherms, due to the small difference between the equivalent moisture contents at mentioned RH a single value representing the moisture content of the whole beam is considered. The initial moisture content at 50\% RH is estimated as 10\% and the new equilibrium moisture level at 90\% RH is approximated as 21\%. They are taken as surface conditions for the moisture transport simulation. Note that these moisture levels are taken as the mean values of adsorption and desorption curves for simplicity. The time variation of RH or equivalently moisture content is expressed by means of the positive part of the sine function, which varies between 10\% to 21\% for the duration of one year (365 days). 

\begin{figure*}[htpb]
     \begin{center}
	\includegraphics[scale=0.60]{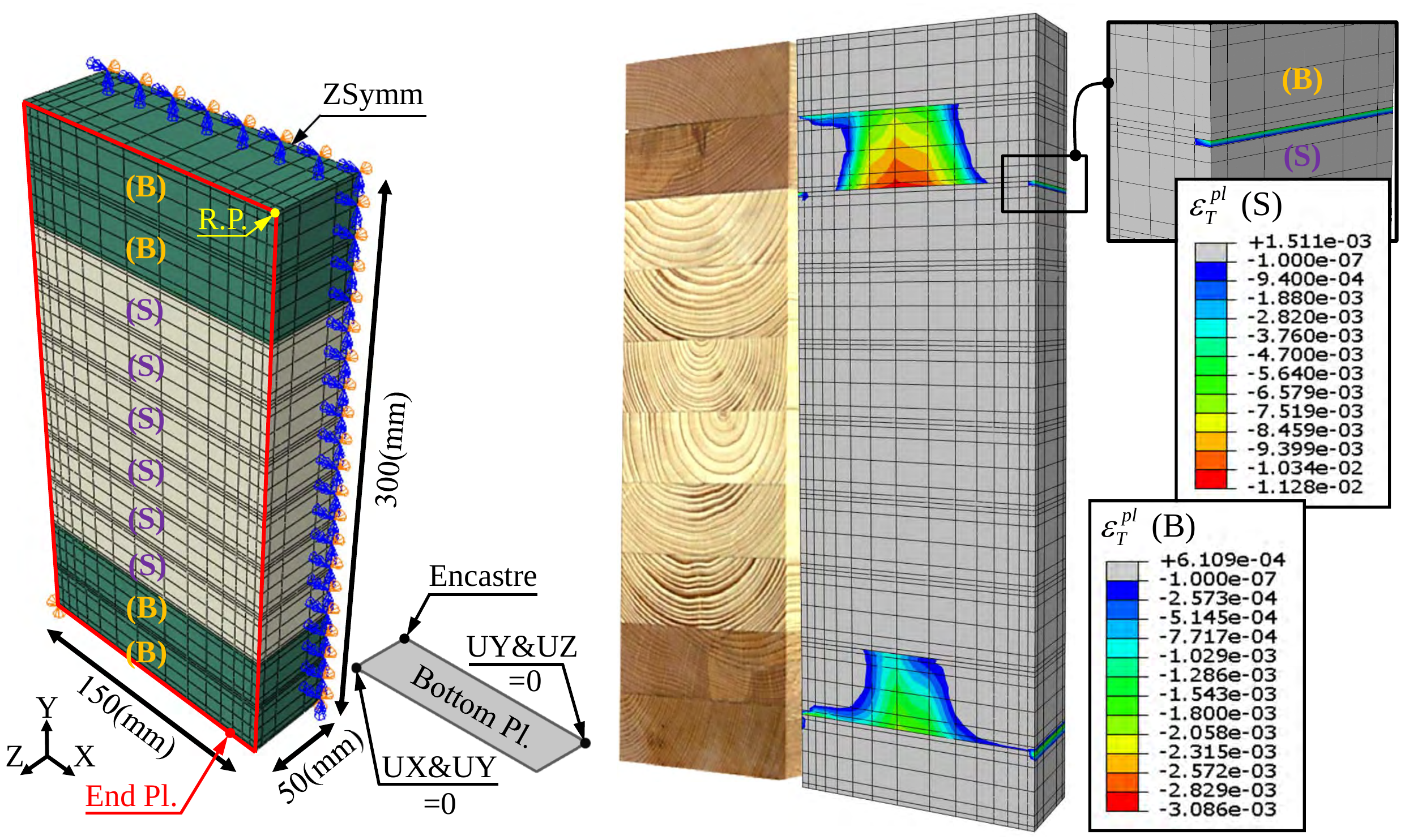}            
   \caption{Left: Geometrical dimensions and finite element discretization of the multi-species beam. (B) and (S) designate beech and spruce, respectively. Right: Distribution of tangential plastic strain after one year ($t=$365 days).}
   \label{fig:HybBeam_11}
   \end{center}
\end{figure*}

In the simulation, two distinct constitutive models consisting of five deformation mechanisms each, related to every wood type are considered. In practice, two material (UMAT) subroutines are defined simultaneously. The adhesive layer between two lamellae is neglected and lamellae are considered as different sections of one part, but with distinct material type and individual cylindrical local material coordinate systems, each. Shared nodes therefore have only one value for the moisture content and adhesive bond lines are fully permeable. As illustrated in Fig.~\ref{fig:HybBeam_11} (left), the simulated system represents a part of a hybrid glulam beam consisting of ten 30mm thick and 150mm wide lamellae, all aligned in the longitudinal direction with the top and bottom two made of beech, while the core is made of six spruce lamellae. In order to avoid modeling the entire length of the beam, a section with the length of 50mm under assumption of generalized plane strain state has been considered. For this purpose, at the top right corner of the cross-section a reference point (R.P.) is defined and the movements of the end plane with red boundaries, i.e., translation along Z-direction and rotation around X and Y axes are kinematically coupled to the reference point (see Fig.~\ref{fig:HybBeam_11} (left)). 

In Fig.~\ref{fig:HybBeam_11} (right) the distribution of the plastic deformations along tangential direction is given. It should be noted that plastic strains in each lamella are illustrated based on the corresponding local material coordinate system. Depending on the position of the center and the orientation of the assigned local cylindrical coordinate systems, the distribution of the plastic strain varies. As it's expected and could also be observed in the calculations, the magnitudes of plastic strain in the beech layers are smaller than the counterpart values in the spruce lamellae. Additionally, the plastic deformations are mostly generated near the adhesive bond-lines. This is due to the occurrence of higher stress concentration in the interface of two adjacent lamellae, resulting from different material properties and material orientations. These stress concentrations, in forms of tension or shear in particular, can lead to the initiation of interfacial cracks that propagate under applied service mechanical loadings and changing environmental conditions. Subsequently, structural integrity and load-carrying capacity are diminished. So, plasticity induced de-bonding could be considered as a critical failure mode of such laminated structural elements, the consequences of which have to be taken into account during design stage. By application of the annual moisture profile during some consecutive years, the role of both short- and long-term responses in the mechanical performance of the hybrid element in addition to the influence of moisture cycles on the development of time- and history-dependent stress states are captured.
\section{Conclusions}\label{Sec:Conclusions}
In the current study, a comprehensive rheological model for the mechanical behavior of wood under simultaneous mechanical loadings and varying environmental conditions is introduced. The considered deformation modes are both short- and long-term responses, namely elastic, plastic, hygro-expansion, visco-elastic creep, and mechano-sorption. To characterize the irrecoverable part of the material response, described as plastic deformation, a general multi-surface plasticity model with three independent failure mechanisms - each one belonging to an anatomical direction under compression - is employed. Moreover, the time-dependent behaviors like visco-elastic and mechano-sorptive creep are simulated through serial association of \emph{Kelvin-Voigt} elements. In order to consider the inherent hygroscopic character of wood in the assessment of timber structures, all material input parameters are defined as function of the moisture level. The numerical integration algorithm along with the utilized approach for the stress update scheme are established on an iteratively additive decomposition of the total strain. All corresponding rheological formulations were developed and implemented into a user-material subroutine (UMAT) within the environment of the finite element method. 

The material model was applied to two important wood species, namely European beech and Norwegian spruce. However due to scarce availability of stress-strain relationships under different loading cases and moisture contents, for beech in particular, some property adaptation procedures had to be implemented to obtain a full set of material data. The functionality and performance of the model and its implementation was evaluated through several examples for uni- and multi-axial loading, as well as for restrained swelling. The examples demonstrated the applicability of the presented material model under any optional load combination, specifically when more than one failure mechanism is activated. In addition, the validation of the presented material model was carried out by comparison with experimental results in a glued-laminated specimen. 
  
The introduced material model is numerically implemented in a general fashion and can also be used for other wood species. By application of moisture-dependent material properties, the rheological model can easily be adapted to predict the mechanical response of different wood types. In this respect, the flexibility and efficiency of the material model has been proven by simulating a hybrid beech/spruce glulam element. In general, the proposed wood constitutive material model can be employed for finite element analysis of moisture-induced stresses in adhesively bonded wood elements. Additionally, this model can be considered as a suitable basis for fracture mechanical analyses or can be combined with cohesive interface elements to predict the initiation and propagation of interfacial cracks. This way we achieve the goal of attaining more realistic and reliable predictions on mechanical performance of wood elements under any arbitrary combinations of loading and moisture content to increase safety and long-term reliability of engineered wood structures.
\section{Acknowledgments}\label{Sec:Acknowledge}
This work was funded by the Swiss National Science Foundation in the National Research Programme NRP 66 - Resource Wood under grant Nr. 406640- 140002: Reliable timber and innovative wood products for structures. We also would like to thank Prof. P. Niemz and Mr. S. Ammann for fruitful discussions.
\begin{appendix}
\counterwithin{equation}{section}
\counterwithin{table}{section}
\counterwithin{figure}{section}
\section{Algorithmic tangent operator for the whole model}\label{Sec:TangentOpt}
\paragraph{Purely elastic or elasto-plastic tangent operator:} From a rheological point of view, the tangent operator corresponding to the first two elements of the constitutive material model (Fig.~\ref{fig:MatModel}), namely the elastic spring and the plastic \emph{Kelvin} element are correlated. In the absence of plastic strain, the plastic \emph{Kelvin} element will be inactive without giving a contribution to the tangent operator. Trivially, the single Jacobian is equal to the elastic stiffness tensor~$\boldsymbol{\mathrm{C}}_{n+1}^{el}=\boldsymbol{\mathrm{C}}_{0,n+1}$. However in the presence of plastic deformation, both elastic and plastic elements contribute to the definition of the elasto-plastic algorithmic tangent modulus \cite{SimoHughes98}:
\begin{equation}
     \label{eq:Cep}
\boldsymbol{\mathrm{C}}_{n+1}^{ep}=\boldsymbol{\Xi}_{n+1}-\sum_{\beta\in\mathrm{\mathbb{S}}_{act}}\sum_{\alpha\in\mathrm{\mathbb{S}}_{act}} \left(G_{n+1}^{\beta\alpha}~\boldsymbol{N}_{\beta,n+1}\otimes\boldsymbol{N}_{\alpha,n+1}\right).
\end{equation}
Note that~$\otimes$ symbolizes the dyadic product of two vectors and~$\boldsymbol{\Xi}_{n+1}$ is the \emph{algorithmic modulus} and is given by the following relationship:
\begin{equation}
     \label{eq:AlgorMod}
\boldsymbol{\Xi}_{n+1}=\left[\boldsymbol{\mathrm{C}}_{0,n+1}^{-1}+\sum_{l=1}^{r}\Delta{\gamma}^{l}_{n+1}
\partial_{\boldsymbol{\sigma}\boldsymbol{\sigma}}^{2}f_{l}\left(\boldsymbol{\sigma}_{n+1},\alpha_{l,n+1},\omega_{n+1}\right)\right]^{-1}=\left[\boldsymbol{\mathrm{C}}_{0,n+1}^{-1}+2\sum_{l=1}^{r} \Delta{\gamma}^{l}_{n+1}\boldsymbol{\mathrm{b}}_{l,n+1}\right]^{-1},
\end{equation}
and
\begin{equation}
     \label{eq:NBeta}
\boldsymbol{N}_{\alpha,n+1}=\boldsymbol{\Xi}_{n+1}:\partial_{\boldsymbol{\sigma}}f_{\alpha,n+1},
\end{equation}
where~$r$ is the number of active yield surfaces and~$G_{n+1}^{\beta\alpha}$ is defined by the succeeding expression:
\begin{equation}
     \label{eq:Gab}
     G_{n+1}^{\beta\alpha} = \left[\partial_{\boldsymbol{\sigma}}f_{\beta,n+1}:\boldsymbol{\Xi}_{n+1}:\partial_{\boldsymbol{\sigma}}f_{\alpha,n+1}+\partial_{\boldsymbol{q}}f_{\beta,n+1}~\mathrm{H}^{\beta\alpha}_{n+1}~\partial_{\boldsymbol{q}}f_{\alpha,n+1}\right]^{-1}.
\end{equation}
Here~$\mathrm{H}^{\beta\alpha}_{n+1}$ denotes the~$\beta\alpha$ component of the matrix of hardening moduli~$\boldsymbol{\mathrm{H}}_{n+1}$ filled with the respective values for every active yield criterion~$\left(H_{l}={\partial{q_{l}}}/{\partial{\alpha_{l}}},~l=\text{R, T, L}\right)$. Eq.~(\ref{eq:Cep}) represents the desired description for algorithmic or the equivalently consistent tangent operator which ensures the quadratic rate of asymptotic convergence of the global iteration. Note that the application of the continuum elasto-plastic moduli yields at most a linear rate of convergence \cite{SimoTaylor85,SimoHughes98,HanHelnII2003}.
\paragraph{Visco-elastic tangent operator:} The analytical solution of Eq.~(\ref{eq:VElasSol}) under assumption of linear variation of stress during time increment~$(\Delta{t}=t_{n+1}-t_{n})$ holds as: \cite{HanHelnII2003}
\begin{equation}
     \label{eq:VisSol}
\boldsymbol{\varepsilon}_{i,n+1}^{ve}=\boldsymbol{\varepsilon}_{i,n}^{ve}~\mathrm{exp}\left(-\frac{\Delta{t}}{\tau_{i}}\right)+\mathrm{\mathbb{T}}_{n}^{ve}\left(\frac{\Delta{t}}{\tau_{i}}\right){\bf C}_{i,n}^{-1}:\boldsymbol{\sigma}_{n}+\mathrm{\mathbb{T}}_{n+1}^{ve}\left(\frac{\Delta{t}}{\tau_{i}}\right){\bf C}_{i,n+1}^{-1}:\boldsymbol{\sigma}_{n+1}.         
\end{equation}
The time functions~$\mathrm{\mathbb{T}}_{n+1}^{ve}\left(\xi_{i}\right)$ and~$\mathrm{\mathbb{T}}_{n}^{ve}\left(\xi_{i}\right)$ in Eq.~(\ref{eq:VisSol}), where ~$\xi_{i}=\Delta{t}/\tau_{i}$, are described as:
\begin{equation}
     \label{eq:TnVe}
\mathrm{\mathbb{T}}_{n+1}^{ve}\left(\xi_{i}\right)=1-\frac{1}{\xi_{i}}\left(1-\mathrm{exp}\left(-\xi_{i}\right)\right), \qquad \mathrm{\mathbb{T}}_{n}^{ve}\left(\xi_{i}\right)=1-\mathrm{exp}\left(-\xi_{i}\right)-\mathrm{\mathbb{T}}_{n+1}^{ve}\left(\xi_{i}\right).   
\end{equation}
The differential form of Eq.~(\ref{eq:VisSol}) after taking the derivative from both sides reads as
\begin{equation}
     \label{eq:DiffVisc}
\mathrm{d}\boldsymbol{\varepsilon}_{i,n+1}^{ve}=\mathrm{\mathbb{T}}_{n+1}^{ve}\left(\frac{\Delta{t}}{\tau_{i}}\right){\bf C}_{i,n+1}^{-1}:\mathrm{d}\boldsymbol{\sigma}_{n+1},          
\end{equation}
that can be rearranged to give the general description of the algorithmic visco-elastic tangent operator~${\bf C}_{i,n+1}^{ve}$:
\begin{equation}
     \label{eq:ViscTangent}
\boldsymbol{\mathrm{C}}_{i,n+1}^{ve}=\boldsymbol{\mathrm{C}}_{i,n+1}/\mathrm{\mathbb{T}}_{n+1}^{ve}\left(\xi_{i}\right)~.
\end{equation}
\paragraph{Mechano-sorptive tangent operator:} Based on an approach similar to the one outlined for visco-elasticity, the response of Eq.~(\ref{eq:MechSol}) can be stated by the following relationship:
\begin{equation}
     \label{eq:MsSol}
\boldsymbol{\varepsilon}_{j,n+1}^{ms}=\boldsymbol{\varepsilon}_{j,n}^{ms}\mathrm{exp}\left(-\frac{\left|\Delta{\omega}\right|}{\mu_{j}}\right)+\mathrm{\mathbb{T}}_{n}^{ms}\left(\frac{\left|\Delta{\omega}\right|}{\mu_{j}}\right){\bf C}_{j,n}^{-1}:\boldsymbol{\sigma}_{n}+\mathrm{\mathbb{T}}_{n+1}^{ms}\left(\frac{\left|\Delta{\omega}\right|}{\mu_{j}}\right){\bf C}_{j,n+1}^{-1}:\boldsymbol{\sigma}_{n+1}.          
\end{equation}
Analogous to Eq.~(\ref{eq:TnVe}) the moisture functions~$\mathrm{\mathbb{T}}_{n+1}^{ms}\left(\xi_{j}\right)$ and~$\mathrm{\mathbb{T}}_{n}^{ms}\left(\xi_{j}\right)$, where~$\xi_{j}=\left|\Delta{\omega}\right|/\mu_{j}$, can be given by the following expressions:
\begin{equation}
     \label{eq:TnMs}
\mathrm{\mathbb{T}}_{n+1}^{ms}\left(\xi_{j}\right)=1-\frac{1}{\xi_{j}}\left(1-\mathrm{exp}\left(-\xi_{j}\right)\right), \qquad \mathrm{\mathbb{T}}_{n}^{ms}\left(\xi_{j}\right)=1-\mathrm{exp}\left(-\xi_{j}\right)-\mathrm{\mathbb{T}}_{n+1}^{ms}\left(\xi_{j}\right).   
\end{equation}
Similar to the approach taken for the visco-elastic Jacobian, differentiating both sides of Eq.~(\ref{eq:MsSol}) and comparing to the standard definition of the tangent operator, leads to the algorithmic mechano-sorptive tangent
\begin{equation}
     \label{eq:MechSTangent}
\boldsymbol{\mathrm{C}}_{j,n+1}^{ms}=\boldsymbol{\mathrm{C}}_{j,n+1}/\mathrm{\mathbb{T}}_{n+1}^{ms}\left(\xi_{j}\right)~.
\end{equation}

It should be noted that hygro-expansion has no contribution to the definition of the total operator. Therefore, after specifying all above-mentioned individual algorithmic operators, the mathematical description of the total Jacobian, corresponding to the entire material model~${\bf{C}}_{n+1}^{T}$ takes the following form:       
\begin{equation}
     \label{eq:TotalTanOp}
 \boldsymbol{\mathrm{C}}^{T}_{n+1}= \begin{dcases}
\left(\boldsymbol{\mathrm{C}}_{n+1}^{ep~~-1}+\sum_{i=1}^{n}\boldsymbol{\mathrm{C}}_{i,n+1}^{ve~~-1}+\sum_{j=1}^{m}\boldsymbol{\mathrm{C}}_{j,n+1}^{ms~~-1}\right)^{-1} & \mbox{, if}~   \boldsymbol{\varepsilon}^{pl(k)}_{n+1}\neq0,     \\
\left(\boldsymbol{\mathrm{C}}_{n+1}^{el~~-1}+\sum_{i=1}^{n}\boldsymbol{\mathrm{C}}_{i,n+1}^{ve~~-1}+\sum_{j=1}^{m}\boldsymbol{\mathrm{C}}_{j,n+1}^{ms~~-1}\right)^{-1} & \mbox{, if}~   \boldsymbol{\varepsilon}^{pl(k)}_{n+1}=0.   
  \end{dcases}
\end{equation}
\nomenclature{$\boldsymbol{\mathrm{C}}_{n+1}^{el}/\boldsymbol{\mathrm{C}}_{n+1}^{ep}$}{Elastic/Elasto-plastic tangent operator (at~$t_{n+1}$)}
\nomenclature{$\boldsymbol{\Xi}_{n+1}$}{Algorithmic modulus (at~$t_{n+1}$)}
\nomenclature{$\mathrm{\mathbb{T}}_{n+1}^{ve/ms}, \mathrm{\mathbb{T}}_{n}^{ve/ms}$}{Visco-elastic time/Mechano-sorptive moisture functions}
\nomenclature{${\bf C}_{i/j,n+1}^{ve/ms}$}{$i/j^{th}$ algorithmic visco-elastic/mechano-sorptive operator (at~$t_{n+1}$)}
\nomenclature{${\bf{C}}_{n+1}^{T}$}{Total algorithmic tangent operator (at~$t_{n+1}$)}
\nomenclature{$\boldsymbol{\varepsilon}^{pl(k)}_{n+1}$}{Plastic strain tensor after $\left(k\right)^{th}$ iteration of the strain decomposition algorithm (at~$t_{n+1}$)}
\section{Material properties for European beech and Norwegian spruce}\label{Sec:Tables}
\begin{table*}[h!]
	  \centering
		\label{table:Elascoef}
		\caption{Coefficients for moisture-dependent engineering constants for European beech \protect\cite{Hering11} and Norway spruce     \protect\cite{GerekeThes09,NeuhausThes81}. (Section~\ref{SSSec:Elastic})} 
\footnotesize
  \renewcommand{\arraystretch}{0.90}
		\begin{tabular}{ccccccccccc}     \toprule
 	 & & $E_{R}$ & $E_{T}$ & $E_{L}$ & $G_{RT}$ & $G_{RL}$ & $G_{TL}$ & $\nu_{TR}$ & $\nu_{LR}$ & $\nu_{LT}$\\
   & &[MPa]&[MPa]&[MPa]&[MPa]&[MPa]&[MPa]&$\left[-\right]$&$\left[-\right]$& $\left[-\right]$\\ 
   & &  &  &  &  &  &     & $\left(\times 10^{-3}\right)$ & $\left(\times 10^{-3}\right)$ & $\left(\times 10^{-3}\right)$ \\     \midrule
       
\rowcolor{black!10} Beech  & $b_0$ & 2565.6 & 885.4  & 17136.7  & 667.8  & 1482   & 1100   & 293.3  & 383.0   & 336.8   \\
\rowcolor{black!10}        & $b_1$ &-59.7   &-23.4   &-282.4    &-15.19  &-15.26  &-17.72  &-1.012  &-8.722   &-9.071   \\
\rowcolor{black!20} Spruce & $s_0$ & 999.64 & 506.08 & 12791.75 & 61.33  & 762.8  & 880.75 & 153.4  & 232.0   & 285.7   \\
\rowcolor{black!20}        & $s_1$ & 3.61   & 5.0    & 15.22    &-1.07   & 5.93   & 1.39   & 10.8   &-8.6     &-41.0    \\
\rowcolor{black!20}        & $s_2$ &-2.09   &-1.35   &-9.01     &-0.06   &-1.99   &-1.39   & 0.398  & 2.8784  & 7.1907  \\
\rowcolor{black!20}        & $s_3$ & 0.0467 & 0.0297 & 0.1885   & 0.0017 & 0.0477 & 0.0277 &-0.0191 &-0.07862 &-0.1642  \\     \bottomrule 
		\end{tabular} 
\end{table*}
\begin{table*}[h!]
	  \centering
\label{table:Hardparam} 
\caption{Coefficients for calculation of moisture-dependent hardening stress for European beech \protect\cite{HeringSaft12,Gattlen13}. (Section~\ref{SSSec:Plastic})}
\footnotesize
  \renewcommand{\arraystretch}{0.90}
    \begin{tabular}{c>{\columncolor[gray]{0.9}}c>{\columncolor[gray]{0.9}}c>{\columncolor[gray]{0.9}}c}     \toprule
Direction & \multicolumn{3}{c}{Beech}  \\ 
\hline
\rowcolor{black!0} & $\beta_{0l}$ & $\beta_{1l}$ & $\beta_{2l}$  \\ 
\rowcolor{black!0} $\left[l\right]$ & $\left[\text{MPa}/\%\right]$ & $\left[\text{MPa}\right]$ & $\left[-\right]$  \\ 
\hline
R & -0.1123 & 2.8765 & 55.29  \\ 
T & -0.1411 & 3.2736 & 65.56  \\ 
L &  0.0    & 2.5180 & 95.3  \\     \bottomrule     
\end{tabular}
\end{table*}
\begin{table*}[h!]
	  \centering
    \label{table:Strenparam}
		\caption{Coefficients for calculation of moisture-dependent strength values for European beech \protect\cite{HeringThes11} and Norway spruce \protect\cite{SchKaliMulti06,SaftKalis11}. (Section~\ref{SSSec:Plastic})}
\footnotesize
  \renewcommand{\arraystretch}{0.90}
\begin{tabular}{c>{\columncolor[gray]{0.9}}c>{\columncolor[gray]{0.9}}c>{\columncolor[gray]{0.8}}c>{\columncolor[gray]{0.8}}c}  \toprule
Strength  & \multicolumn{2}{c}{Beech} & \multicolumn{2}{c}{Spruce} \\
\hline
\rowcolor{black!0} & ~$z_{b0}$~  & ~$z_{b1} $~      & ~$z_{s0}$~  & ~$z_{s1}$~    \\
\rowcolor{black!0} & ~[MPa/\%]~  & ~[MPa]~          & ~[MPa/\%]~  & ~[MPa]~       \\
\hline
    $f_{c,R} $~    & -0.5789     & 20.32   & -0.270   &  9.24                    \\
    $f_{c,T} $~    & -0.2084     & 7.965   & -0.270   &  9.24                    \\
    $f_{c,L} $~    & -3.5040     & 103.9   & -2.150   &  68.8                    \\    
    $f_{t,L} $~    & -1.4650     & 106.9   & -3.275   & 104.8                    \\
    $f_{s,RT}$~    & -0.1213     & 4.982   & -0.063   & 2.585                    \\ 
    $f_{s,RL}$~    & -0.3884     & 15.51   & -0.227   & 9.064                    \\ 
    $f_{s,TL}$~    & -0.3861     & 16.21   & -0.178   & 7.477                    \\     \bottomrule
		\end{tabular}
\end{table*}
\begin{table*}[h!]
	  \centering
    \label{table:Hygcoef}
		\caption{Swelling/shrinkage coefficients for European beech \protect\cite{HeringThes11} and Norway spruce \protect\cite{GerekeThes09,NeuhausThes81}. (Section~\ref{SSSec:Hygro})}
\footnotesize
  \renewcommand{\arraystretch}{0.90}
		\begin{tabular}{cccc}
     \toprule
			Hygro-expansion coefficient & ~$\alpha_{R}$ $\left[1/\%\right]$~ & ~$\alpha_{T}$ $\left[1/\%\right]$~ & ~$\alpha_{L}$ $\left[1/\%\right]$~\\
         \hline
\rowcolor{black!10}    Beech & 0.00191 & 0.00462 & 0.00011 \\
\rowcolor{black!20}    Spruce  & 0.00170 & 0.00330 & 0.00005 \\     \bottomrule
		\end{tabular}
\end{table*}
\begin{table*}[h!]
	  \centering
    \label{table:Gamval}
		\caption{Coefficients for calculation of moisture- and time-dependent entry of the visco-elastic compliance tensor parallel to grain for European beech \protect\cite{HeringCreep12} and dimensionless scalar parameters for Norway spruce \protect\cite{Fortino09}. (Section~\ref{SSSec:Visco})}
\footnotesize
      \renewcommand{\arraystretch}{0.90}     
\begin{tabular}{>{\columncolor[gray]{0.9}}c>{\columncolor[gray]{0.9}}c>{\columncolor[gray]{0.9}}c>{\columncolor[gray]{0.9}}c>{\columncolor[gray]{0.8}}c>{\columncolor[gray]{0.8}}c>{\columncolor[gray]{0.8}}c}
		     \toprule
\multicolumn{4}{c}{Beech} & \multicolumn{3}{c}{Spruce} \\ 
         \hline
\rowcolor{black!0}~$i [-]$~ & ~$J_{i1}\left[\text{MPa}^{-1}\right]$~ & ~$J_{i0}\left[\text{MPa}^{-1}\right]$~ & ~$\tau_{i}\left[\text{h}\right]$~  & ~$i [-]$~ & ~$\gamma_{i}^{ve}$~ & ~$\tau_{i}\left[\text{h}\right]$~ \\
         \hline
        1 & 1.11e-6  & -4.99e-6 & 0.82     & 1    &  $1/0.085$  &  2.4 \\
        2 & 9.84e-7  & -5.22e-6 & 60.86    & 2    &  $1/0.035$  &  24 \\
        3 & 8.51e-7  & -5.97e-6 & 8.90     & 3    &  $1/0.070$  &  240 \\
        4 & 2.82e-6  &  1.03e-5 & 3427.65  & 4    &  $1/0.200$  &  2400\\     \bottomrule
		\end{tabular}
\end{table*}
\begin{table*}[h!]
	  \centering
  	\caption{Longitudinal and tangential components of the mechano-sorptive compliance tensor for European beech and Norway spruce \protect\cite{Fortino09}. (Section~\ref{SSSec:Mechano})}
    \label{table:Mechprops}		
\footnotesize
     	  \renewcommand{\arraystretch}{0.90}
\begin{tabular}{>{\columncolor[gray]{0.9}}c>{\columncolor[gray]{0.9}}c>{\columncolor[gray]{0.9}}c>{\columncolor[gray]{0.9}}c>{\columncolor[gray]{0.8}}c>{\columncolor[gray]{0.8}}c>{\columncolor[gray]{0.8}}c>{\columncolor[gray]{0.8}}c}
		     \toprule
		     \multicolumn{4}{c}{Beech} & \multicolumn{4}{c}{Spruce} \\ 
		              \hline
\rowcolor{black!0}  $ j$ & $J_{jT0}^{ms}\left[\text{MPa}^{-1}\right]$ & $J_{jL}^{ms}\left[\text{MPa}^{-1}\right]$ & $\mu_{j}\left[-\right]$ & $j$ & $J_{jT0}^{ms}\left[\text{MPa}^{-1}\right]$ & $J_{jL}^{ms}\left[\text{MPa}^{-1}\right]$ & $\mu_{j}\left[-\right]$\\
         \hline
        1 & 0.0004 & $\frac{0.1142}{E_{L}}$ & 1    &  1  &  0.0006  & $\frac{0.175}{E_{L}}$  & 1     \\
        2 & 0.0004 & $\frac{0.3200}{E_{L}}$ & 10   &  2  &  0.0006  & $\frac{0.490}{E_{L}}$  & 10    \\
        3 & 0.0033 & $\frac{0.0228}{E_{L}}$ & 100  &  3  &  0.0050  & $\frac{0.035}{E_{L}}$  & 100   \\
         \bottomrule
		\end{tabular}
\end{table*}
\begin{table*}[h!]
	  \centering
		\caption{Parameters for calculation of diffusion coefficients as an exponential function of moisture content for European beech \protect\cite{HeringThes11} and Norway spruce \protect\cite{SaftKalis11}. (Section~\ref{SSec:Moisture})}
    \footnotesize     	
	  \renewcommand{\arraystretch}{0.90}	\begin{tabular}{>{\columncolor[gray]{0.9}}c>{\columncolor[gray]{0.9}}c>{\columncolor[gray]{0.9}}c>{\columncolor[gray]{0.8}}c>{\columncolor[gray]{0.8}}c>{\columncolor[gray]{0.8}}c}
		     \toprule
		     		     \multicolumn{3}{c}{Beech} & \multicolumn{3}{c}{Spruce} \\ 
		              \hline
\rowcolor{black!0}			  Direction~ & ~$\mathrm{D}_{0l}$~    &  ~$\alpha_{0l}$~ & Direction~ & ~$\mathrm{D}_{0l}$~    &  ~$\alpha_{0l}$~         \\
\rowcolor{black!0}        $[l]$~     & ~$[\frac{mm^{2}}{h}]$~ &  ~$[-]$~         & $[l]$~     & ~$[\frac{mm^{2}}{h}]$~ &  ~$[-]$~                \\ 
         \hline
         Radial       &  0.02630 &  0.199724  &  Radial       & 0.288 & 4   \\
         Tangential   &  0.00370 &  0.265280  &  Tangential   & 0.288 & 4   \\
         Longitudinal &  21.8999 &  -0.038545 &  Longitudinal & 0.720 & 4   \\
         \bottomrule
		\end{tabular}
    \label{table:Moisprops}
\end{table*}
\begin{table*}[h!]
	  \centering
		\caption{Residual norm values of iterations of the \emph{Newton} solution procedure for some typical increments for the time between 62.5h and 65h in Example 1 (Section~\ref{Sec:Verif}). The control parameter for the convergence criterion is taken as~$R^{\alpha}=5\times10^{-3}$.}  
\footnotesize
  \renewcommand{\arraystretch}{0.90}
\begin{tabular}{cccccccc}  \toprule
     Iteration& \multicolumn{7}{c}{Residual norm}                  \\
\hline     
         \#   & Inc. 429   & Inc. 445   &  Inc. 448   & Inc. 450   &  Inc. 454   & Inc. 457   & Inc. 464     \\
\hline
			1       & 2.9128e-02 & 3.7130e-01 &  3.6250e-01 & 3.4960e-01 &  4.0181e-02 & 7.2910e-03 & 3.8530e-01   \\ 
			2       & 2.3883e-04 & 1.1859e-02 &  1.1477e-02 & 1.0810e-02 &  1.5068e-14 & 2.8608e-14 & 1.3920e-02   \\
			3       & -          & 5.8554e-05 &  7.0931e-05 & 1.4326e-04 &  -          & -          & 1.7164e-04   \\
    \bottomrule
		\end{tabular}
    \label{table:ConverNorm}
\end{table*}
\begin{table*}[h!]
	  \centering
		\caption{Moisture-dependent material properties for PRF adhesive \protect\cite{OliverMech13,AdhCME,AdhDiffus} (Section~\ref{Sec:Verif}). $E_{adh}$,~$\alpha_{adh}$, and~${D}_{adh}$ designate Young's modulus, coefficient of moisture expansion, and diffusion coefficient, respectively. PRF Poisson's ratio is assumed as a constant value equal to~$\nu_{adh}=0.3$.}
\footnotesize
  \renewcommand{\arraystretch}{0.90}
		\begin{tabular}{cccc}
     \toprule
      \multicolumn{2}{c}{$E_{adh}(\omega)=a_0+a_1\omega+a_2\omega^{2}+a_3\omega^{3}$}                                 \\  
									$a_0$~$\mathrm{[MPa]}$~&$a_1$~$\mathrm{[MPa/\%]}$~&$a_2$~$\mathrm{[MPa/{\%}^2]}$~&$a_3$~$\mathrm{[MPa/{\%}^3]}$    \\ 
									 4176       &-176.9         &19.38              &-0.8521               \\
         \hline 
                  $\mathrm{\alpha_{adh}}$~$\left[1/\%\right]$~& 0.00172869 &     &                \\
         \hline 
                  ${D}_{adh}(\omega)=D_{0}~(e^{\alpha_{0}\omega})$&     &     &                                     \\
									$D_{0}$~$[\frac{mm^{2}}{h}]$~& $4.04736$$\mathrm{\times}$$10^{-4}$  &~$\mathrm{\alpha_{0}}$~$[-]$~& $0.231$               \\
    \bottomrule
		\end{tabular}
    \label{table:AdhPRF}
\end{table*}
\begin{table*}[h!]
	  \centering
		\caption{Comparisons of experimental and numerical deformations in the glued-laminated sample after the~$1^{st}$ and~$2^{nd}$ drying steps at ~($t=792\mathrm{h}$) and~($t=3024\mathrm{h}$), respectively (Experimental validation-Section~\ref{Sec:Verif}). The subscripts (exp) and (num) refer to the experimental and numerical values.}  
\footnotesize
  \renewcommand{\arraystretch}{0.90}
\begin{tabular}{cccccc}  \toprule
          & ~$\mathrm{H_{exp}/H_{num}}$~ & ~$\mathrm{h_{exp}/h_{num}}$~ & ~$\mathrm{d_{exp}/d_{num}}$~ & ~$\mathrm{W_{exp}/W_{num}}$~ & ~$\mathrm{L_{exp}/L_{num}}$  \\
          & ~[mm]~ & ~[mm]~ & ~[mm]~ & ~[mm]~ & ~[mm]       \\
\hline
$\mathrm{1^{st} drying}$~ & 9.39/9.07 & 1.83/1.58 & 7.67/7.50 & 97.19/95.56 & 100.54/99.77                    \\ 
\hline
$\mathrm{2^{nd} drying}$~ & 9.48/9.15 & 1.85/1.67 & 7.67/7.50 & 96.36/95.53 & 100.26/99.74                    \\
    \bottomrule
		\end{tabular}
    \label{table:PRFExpNum}
\end{table*}
\end{appendix}
\newpage
\begin{thenomenclature} 

 \nomgroup{A}

  \item [{$\alpha_{l,n+1}^{(p)}$}]\begingroup Internal hardening variable after~$\left(p\right)^{th}$ iteration of return mapping (at~$t_{n+1}$),~$l=$R, T, L\nomeqref {62}
		\nompageref{18}
  \item [{$\alpha_{l,n+1}^{(Trial)}$}]\begingroup Trial internal hardening variable (at~$t_{n+1}$),~$l=$R, T, L\nomeqref {62}
		\nompageref{18}
  \item [{$\boldsymbol{\alpha}_{\omega}$}]\begingroup Hygro-expansion coefficients tensor\nomeqref {18}
		\nompageref{10}
  \item [{$\boldsymbol{\mathrm{a}}_{l}, \boldsymbol{\mathrm{b}}_{l}$}]\begingroup Strength tensors\nomeqref {8}
		\nompageref{7}
  \item [{$\boldsymbol{\mathrm{C}}_{n+1}^{el}/\boldsymbol{\mathrm{C}}_{n+1}^{ep}$}]\begingroup Elastic/Elasto-plastic tangent operator (at~$t_{n+1}$)\nomeqref {A.12}
		\nompageref{29}
  \item [{$\boldsymbol{\mathrm{R}}^{el}, \boldsymbol{\mathrm{R}}_{i/j}^{ve/ms}$}]\begingroup Elastic,~$i/j^{th}$ visco-elastic/mechano-sorptive residual vector\nomeqref {62}
		\nompageref{18}
  \item [{$\boldsymbol{\sigma}/\boldsymbol{\sigma}_{i}^{ve}/\boldsymbol{\sigma}_{j}^{ms}$}]\begingroup Total stress/$i^{th}$ visco-elastic stress/$j^{th}$ mechano-sorptive stress tensor\nomeqref {30}
		\nompageref{13}
  \item [{$\boldsymbol{\sigma}_{n+1}^{(p)}$}]\begingroup Total stress tensor after~$\left(p\right)^{th}$ iteration of return-mapping (at~$t_{n+1}$)\nomeqref {62}
		\nompageref{18}
  \item [{$\boldsymbol{\sigma}_{n+1}^{(Trial)}$}]\begingroup Trial total stress tensor (at~$t_{n+1}$)\nomeqref {62}
		\nompageref{18}
  \item [{$\boldsymbol{\varepsilon}/\boldsymbol{\varepsilon}^{el}/\boldsymbol{\varepsilon}^{pl}/\boldsymbol{\varepsilon}^{\omega}$}]\begingroup Total strain/Elastic strain/Plastic strain/Hygro-expansion strain tensor\nomeqref {3}
		\nompageref{4}
  \item [{$\boldsymbol{\varepsilon}^{el(p)/pl(p)}_{n+1}$}]\begingroup Elastic/Plastic strain tensor after~$\left(p\right)^{th}$ iteration of return-mapping (at~$t_{n+1}$)\nomeqref {62}
		\nompageref{18}
  \item [{$\boldsymbol{\varepsilon}^{el(Trial)/pl(Trial)}_{n+1}$}]\begingroup Trial elastic/Trial plastic strain tensor (at~$t_{n+1}$)\nomeqref {62}
		\nompageref{18}
  \item [{$\boldsymbol{\varepsilon}^{pl(k)}_{n+1}$}]\begingroup Plastic strain tensor after $\left(k\right)^{th}$ iteration of the strain decomposition algorithm (at~$t_{n+1}$)\nomeqref {A.12}
		\nompageref{29}
  \item [{$\boldsymbol{\varepsilon}_{i/j,n+1}^{ve/ms(k)}$}]\begingroup $i/j^{th}$ visco-elastic/mechano-sorptive strain tensor after $\left(k\right)^{th}$ iteration of the strain decomposition algorithm (at~$t_{n+1}$)\nomeqref {62}
		\nompageref{18}
  \item [{$\boldsymbol{\varepsilon}_{i/j}^{ve/ms}$}]\begingroup $i/j^{th}$ visco-elastic/mechano-sorptive strain tensor\nomeqref {3}
		\nompageref{4}
  \item [{$\boldsymbol{\Xi}_{n+1}$}]\begingroup Algorithmic modulus (at~$t_{n+1}$)\nomeqref {A.12}
		\nompageref{29}
  \item [{$\boldsymbol{R}_{n+1}^{pl(p)}$}]\begingroup Plastic residual vector after~$\left(p\right)^{th}$ iteration of return-mapping (at~$t_{n+1}$)\nomeqref {62}
		\nompageref{18}
  \item [{$\Delta\boldsymbol{\sigma}_{n+1}^{(k)}$}]\begingroup Change of the total stress tensor after~$\left(k\right)^{th}$ iteration\nomeqref {62}
		\nompageref{18}
  \item [{$\Delta\boldsymbol{\varepsilon}_{i/j,n+1}^{ve/ms(k)}$}]\begingroup Change of the~$i/j^{th}$ visco-elastic/mechano-sorptive strain tensor after~$k^{th}$ iteration\nomeqref {62}
		\nompageref{18}
  \item [{$\Delta\boldsymbol{\varepsilon}_{i/j,n+1}^{ve/ms}$}]\begingroup Increment of~$i/j^{th}$ visco-elastic/mechano-sorptive strain tensor from~$t_{n}$ to~$t_{n+1}$\nomeqref {62}
		\nompageref{18}
  \item [{$\Delta{\alpha_{l,n+1}}$}]\begingroup Increment of the internal hardening variable from~$t_{n}$ to~$t_{n+1}$,~$l=$R, T, L\nomeqref {62}
		\nompageref{18}
  \item [{$\Delta{\boldsymbol{\varepsilon}^{pl(p)}_{n+1}}/\Delta{\alpha_{l,n+1}^{(p)}}$}]\begingroup Change of the plastic strain tensor/Change of the internal hardening variable after~$\left(p\right)^{th}$ iteration of return-mapping\nomeqref {62}
		\nompageref{18}
  \item [{$\Delta{\boldsymbol{\varepsilon}^{pl}_{n+1}}$}]\begingroup Increment of the plastic strain tensor from~$t_{n}$ to~$t_{n+1}$\nomeqref {62}
		\nompageref{18}
  \item [{$\Delta{\gamma}^{l(p)}_{n+1}/\delta\Delta{\gamma}^{l(p)}_{n+1}$}]\begingroup Non-rate form/Change of the consistency parameter after~$\left(p\right)^{th}$ iteration of return-mapping,~$l=$R, T, L\nomeqref {62}
		\nompageref{18}
  \item [{$\Delta{t}/t_{n+1},t_{n}$}]\begingroup Time step/Time moments~$n$ and~$n+1$\nomeqref {30}
		\nompageref{13}
  \item [{$\dot{\gamma}^{l}$}]\begingroup Rate form of the plastic consistency parameter,~$l=$R, T, L\nomeqref {12}
		\nompageref{9}
  \item [{$\gamma_{i}^{ve}/\gamma_{j}^{ms}$}]\begingroup $i^{th}$ scalar visco-elastic/$j^{th}$ scalar mechano-sorptive compliance fraction\nomeqref {30}
		\nompageref{13}
  \item [{$\mathrm{\mathbb{S}}_{adm/act}$}]\begingroup Set of admissible/active yield constraints\nomeqref {12}
		\nompageref{9}
  \item [{$\mathrm{\mathbb{T}}_{n+1}^{ve/ms}, \mathrm{\mathbb{T}}_{n}^{ve/ms}$}]\begingroup Visco-elastic time/Mechano-sorptive moisture functions\nomeqref {A.12}
		\nompageref{29}
  \item [{$\omega/\omega_{FS}/\omega_{\textit{0}}$}]\begingroup Current moisture content/Fiber saturation/Initial reference moisture level in \%\nomeqref {18}
		\nompageref{10}
  \item [{$\partial_{\boldsymbol{\sigma}}f_{l}$}]\begingroup Plastic flow direction tensor,~$l=$R, T, L\nomeqref {12}
		\nompageref{9}
  \item [{$\partial_{\boldsymbol{q}}f_{l}$}]\begingroup Hardening strain flow direction\nomeqref {62}
		\nompageref{18}
  \item [{$\psi/\phi(T,\omega)$}]\begingroup Free energy/Thermal energy function\nomeqref {3}
		\nompageref{4}
  \item [{$\psi^{el/ve/ms}$}]\begingroup Elastic/Visco-elastic/Mechano-sorptive strain energy\nomeqref {3}
		\nompageref{4}
  \item [{$\tau_{i}/\mu_{j}$}]\begingroup $i^{th}$ characteristic retardation time/$j^{th}$ characteristic moisture\nomeqref {30}
		\nompageref{13}
  \item [{$\text{c}/\rho_{0}/\text{T}/c_{T}$}]\begingroup Water concentration/Oven-dry wood density/Temperature/Specific heat\nomeqref {36}
		\nompageref{13}
  \item [{$\text{{\bf D}}/\text{{\bf K}}$}]\begingroup Tensor of diffusion coefficients/thermal conductivity coefficients\nomeqref {36}
		\nompageref{13}
  \item [{$f_{c,l}$}]\begingroup Normal compressive strength of the material,~$l=$R, T, L\nomeqref {8}
		\nompageref{7}
  \item [{$f_{l,n+1}^{(p)}/q_{l,n+1}^{(p)}$}]\begingroup Yield function/Hardening function value after~$\left(p\right)^{th}$ iteration (at~$t_{n+1}$),~$l=$R, T, L\nomeqref {62}
		\nompageref{18}
  \item [{$f_{l,n+1}^{(Trial)}/q_{l,n+1}^{(Trial)}$}]\begingroup Trial yield function/Trial hardening function value (at~$t_{n+1}$),~$l=$R, T, L\nomeqref {62}
		\nompageref{18}
  \item [{$f_{l}$}]\begingroup Yield function,~$l=$R, T, L\nomeqref {8}
		\nompageref{7}
  \item [{$f_{s,RT/RL/TL}$}]\begingroup Shear strength in RT/RL/TL plane\nomeqref {8}
		\nompageref{7}
  \item [{$f_{t,L}$}]\begingroup Normal longitudinal tensile strength\nomeqref {8}
		\nompageref{7}
  \item [{$H_{l}$}]\begingroup Hardening modulus,~$l=$R, T, L\nomeqref {12}
		\nompageref{9}
  \item [{$J_{iL}^{ve}$}]\begingroup Longitudinal component of~$i^{th}$ visco-elastic compliance tensor\nomeqref {30}
		\nompageref{13}
  \item [{$J_{jL/jT0}^{ms}$}]\begingroup Longitudinal/Tangential entry of the~$j^{th}$ mechano-sorptive compliance tensor\nomeqref {30}
		\nompageref{13}
  \item [{$n/m$}]\begingroup Number of visco-elastic/mechano-sorptive \emph{Kelvin-Voigt} elements\nomeqref {3}
		\nompageref{4}
  \item [{$P_{b}/P_{s}$}]\begingroup European beech/Norway spruce property\nomeqref {6}
		\nompageref{5}
  \item [{$q_{l}$}]\begingroup Plastic hardening function,~$l=$R, T, L\nomeqref {8}
		\nompageref{7}
  \item [{$r/r_{adm}$}]\begingroup Number of active yield mechanisms/Number of active yield constraints\nomeqref {12}
		\nompageref{9}
  \item [{$R^{\alpha}$}]\begingroup Residual control parameter\nomeqref {62}
		\nompageref{21}
  \item [{${\bf C}_{0}/{\bf C}_{i}/{\bf C}_{j}$}]\begingroup Elastic stiffness/$i^{th}$ visco-elastic stiffness/$j^{th}$ mechano-sorptive stiffness tensor\nomeqref {30}
		\nompageref{13}
  \item [{${\bf C}_{0}^{-1}/{\bf C}_{i}^{-1}/{\bf C}_{j}^{-1}$}]\begingroup Elastic compliance/$i^{th}$ visco-elastic compliance/$j^{th}$ mechano-sorptive compliance tensor\nomeqref {30}
		\nompageref{13}
  \item [{${\bf C}_{i/j,n+1}^{ve/ms}$}]\begingroup $i/j^{th}$ algorithmic visco-elastic/mechano-sorptive operator (at~$t_{n+1}$)\nomeqref {A.12}
		\nompageref{29}
  \item [{${\bf J}_{\omega}/{\bf J}_{T}$}]\begingroup Body moisture/Heat flux vector\nomeqref {36}
		\nompageref{13}
  \item [{${\bf{C}}_{n+1}^{T}$}]\begingroup Total algorithmic tangent operator (at~$t_{n+1}$)\nomeqref {A.12}
		\nompageref{29}
  \item [{${\bf{R}}_{n+1}^{(k+1)}$}]\begingroup Generalized residual vector after~$\left(k+1\right)^{th}$ iteration of the strain decomposition algorithm\nomeqref {62}
		\nompageref{18}

\end{thenomenclature}

\end{document}